\documentstyle[equation,12pt]{article}
\newcommand{\Rrr}{\cal}
\newcommand{\Bbb}{\bf}
\newcommand{\ar}{\alpha}

\newcommand{\dd}{\delta}
\newcommand{\ld}{\lambda}

\newcommand{\be}{\begin{equation}}
\newcommand{\ee}{\end{equation}}
\newcommand{\bea}{\begin{eqnarray}}
\newcommand{\eea}{\end{eqnarray}}
\newcommand{\nn}{\nonumber}

\def\theequation{\arabic{section}.\arabic{equation}}
\newcommand{\bse}{\begin{subequations}}
\newcommand{\ese}{\end{subequations}}
\begin{document}
\begin{flushright}
{\large INS \# 212/92}
\end{flushright}

\begin{center}
{\Large {\bf INTEGRABILITY AND FUSION ALGEBRA }
   \vskip .2cm
{\bf FOR QUANTUM MAPPINGS}}\\
\vspace{.4in}

{\large F.W. Nijhoff}${^\dagger}$\footnote{${^\dagger}$\ E-mail: {\bf
NIJHOFF@SUN.MCS.CLARKSON.EDU}}
\vskip .1in

{\small {\em Department of Mathematics and Computer Science\\
and Institute for Nonlinear Studies,\\
Clarkson University, Potsdam NY 13699-5815, USA}}\\
\vspace{.1in}

{\small and}\\
\vskip .1in

{\large H.W. Capel}\\
\vskip .1in

{\small {\em Institute for Theoretical Physics, University of Amsterdam,\\
Valckenierstraat 65, 1018 XE Amsterdam, The Netherlands}}
\vspace{.1in}

{\small December 1992}\\
\end{center}
\vspace{1in}

\centerline{ {\bf Abstract} }
\vspace{.5in}

We apply the fusion procedure to a quantum Yang-Baxter algebra
associated with time-discrete integrable systems, notably integrable
quantum mappings. We present a general construction of higher-order
quantum invariants for these systems. As an important class of examples,
we present the Yang-Baxter structure of the Gel'fand-Dikii mapping
hierarchy, that we have introduced in previous papers, together with
the corresponding explicit commuting family of quantum invariants.
\vspace{1in}

\addtocounter{footnote}{-1}

\pagebreak

\section{Introduction}

Discrete integrable quantum models, in which the spatial variable is
discretized, have played an important role in the
development of the quantum inverse scattering method \cite{F,KIB},
and subsequently in the advent of quantum groups, cf.
\cite{KR0,Jim}, cf. also \cite{Drin}-\cite{FRT}.
In the modern developments in string theory and conformal field theory
such models also play a particularly important role, cf. e.g.
\cite{FF}-\cite{Smir}. However, quantum models, in which also the
time-flow is discretized (i.e.  whose classical counterparts are
integrable partial difference equations), have not been studied widely
until recently, \cite{NPC}. The study of integrable systems with a
discrete time
evolution is certainly not new. Exactly solvable lattice models in
statistical mechanics form a widely studied class of such systems, the
transfer matrix being the operator that generates the discrete-time
flow. Also integrable partial difference equations
(discrete counterparts of soliton systems) have been studied,
(cf. \cite{NPCQ} and references therein). Recently, these systems have
become of interest in
connection with the construction of exactly integrable {\em mappings},
i.e. finite-dimensional systems with discrete time-flow,
\cite{PNC,CNP}. Their integrability is to
be understood in the sense that the discrete time-flow is the
iterate of a canonical transformation, preserving a suitable
symplectic form, and which carries exact invariants which are in
involution with respect to this symplectic form. Other types of
integrable mappings have been considered also in
the recent literature, (cf. e.g. \cite{Ves} for a review).

Recently, a theory of integrable {\em quantum} mappings was
formulated in \cite{NCP,NC}, cf. also \cite{Exet}. These are the
discrete-time quantum
systems which are obtained from the classical mappings by quantization
within the Yang-Baxter formalism. For the discrete-time systems it
turns out that a beautiful structure arises: it was pointed out in
\cite{NC}, that in contrast to the continuous-time quantum system,
in the case of mappings one has to take into account both spatial
as well as temporal part of the quantum Lax system that governs the
time-evolution. However, the full quantum Yang-Baxter structure
containing both parts carries a consistent set of universal algebraic
relations, reminiscent of the algebraic structure that was proposed
for the cotangent bundle of a quantum group, \cite{AF}.

For the mappings of KdV and MKdV type that were considered in
\cite{NCP} and \cite{Exet}, a construction of exact quantum invariants
is given by expanding a quantum deformation of the trace of the
monodromy matrix of the model. This will yield in principle a large
enough family of commuting quantum invariants to `diagonalize' the
discrete-time model. The deformation, introduced by a quantum
$K$-matrix (in the spirit of ref. \cite{Skly}, cf. also \cite{Cher}),
is established quite
straightforwardly from the relations supplied by the full Yang-Baxter
structure. However, in most examples of quantum mappings, notably those
related to the higher-order members of the Gel'fand-Dikii (GD) hierarchy,
cf. \cite{NPCQ,NC}, one may expect that the trace of the
monodromy matrix  does not yield a large enough family of commuting
invariants. In those cases, one needs the analogues (i.e. the
$K$-deformed versions) of
the higher-order quantum minors and quantum determinants, to supply
the remaining invariants. It is the purpose of this paper to give
a general construction of higher-order quantum invariants of integrable
mappings within the mentioned full Yang-Baxter structure.

The outline of this paper is as follows. In section 2, we briefly
summarize the full (spatial and temporal) Yang-Baxter structure for
mappings. In section 3, we develop the fusion algebra, i.e. the
algebra of higher-order tensor products of generators of the
extended Yang-Baxter algebra. In section 4, we use the fusion
relations to give a construction of exact higher-order quantum
invariants, corresponding to the $K$-deformed quantum minors and
determinants. Furthermore, we give a proof that these will yield
a commuting family of quantum operators. Finally, in section 5, we
apply the general construction to obtain the quantum invariants
of the mappings in the Gel'fand-Dikii hierarchy. In order to make the
paper selfcontained, we supply proofs of most of the basic relations,
even though we have no doubt that some of these must be known to the
experts.

\section{Integrable Quantum Mappings}
\setcounter{equation}{0}

We focuss on the quantization of the mappings that
arise from integrable partial difference equations by finite-dimensional
reductions. Both on the classical as well as on
the quantum level they arise as the compatibility conditions of a
discrete-time ZS (Zakharov-Shabat) system
\be
L'_n(\lambda ) M_n(\lambda )\ =\ M_{n+1}(\lambda ) L_n(\lambda )
\    ,  \label{eq:ZS}
\ee
in which $\lambda$ is a spectral parameter, $L_n$ is the lattice
translation
operator at site $n$, and the prime denotes the discrete time-shift
corresponding to another translation on a two-dimensional lattice. As $L$ and
$M$, in the quantum case, depend on quantum operators (acting on some
well-chosen Hilbert space $\cal H$), the question of operator
ordering becomes important. Throughout we impose
as a normal order the order which is induced by the lattice
enumeration,
with $n$ increasing from the left to the right. Finite-dimensional
mappings are obtained from (\ref{eq:ZS}) imposing a periodicity condition
$L_n(\lambda)=L_{n+P}(\lambda)$, $M_n(\lambda)=M_{n+P}(\lambda)$,
for some period $P$.

We summarize here briefly the (non-ultralocal) Yang-Baxter structure
for integrable quantum mappings, that was introduced in \cite{NC}.

\paragraph{Yang-Baxter structure}

The quantum $L$-operator for the mappings that we consider here
obeys commutation relations of non-ultralocal type, i.e.
\bse  \label{eq:LLL}
\bea
R_{12}^+\,L_{n,1}\,L_{n,2} &=& L_{n,2}\,L_{n,1}\,R_{12}^-\  ,
\label{eq:RLL} \\
L_{n+1,1} S^+_{12}\,L_{n,2} &=& L_{n,2}\,L_{n+1,1}\   ,
\label{eq:LSL} \\
 L_{n,1}\, L_{m,2} &=& L_{m,2}\,L_{n,1} \, , \, \mid n-m \mid
 \geq 2\    , \label{eq:LL}
\eea\ese
with non-trivial commutation relations only between $L$-operators
on the same site and on neighbouring sites.

The notation we use is the following. $R_{ij}^{\pm}$ and
$S_{ij}^{\pm}$ are elements in $End(V_i\otimes V_j)$, where
$V_i$ and $V_j$ are vector spaces, typically the representation
spaces of some quantum group, but here to
be concrete we take them to be different copies of
$V=\Bbb C^N$, and both $R^{\pm}$ and
$S^{\pm}$ in $End(V\otimes V)$. The indices $i,j$ refer to the embedding
$End(V\otimes V)\hookrightarrow End(\otimes_{\alpha} V_{\alpha})$,
in a tensor product of a yet unspecified number of factors.
We adopt the usual convention that the subscripts $1,2,\dots$ denote
factors in a matricial tensor product,
i.e. $A_{i_1,i_2,\dots ,i_M}=A_{i_1,i_2,\dots ,i_M}(\lambda _1,
\lambda _2, \dots ,\lambda _M)$ denotes a matrix acting nontrivially
only on the factors labeled by $i_1,i_2,\dots ,i_M$ of a tensor
product $\otimes_{\ar}\,V_{\ar}$,
of vector spaces $V_{\ar}$ and trivially on the other factors.
Whenever there is no cause for ambiguity, we suppress in the notation
the explicit dependence on the
spectral parameters $\lambda_i$, adopting the convention that each one
accompanies its respective factor in the tensor product. Thus, for
instance, $L_{n,1}$ and $L_{n,2}$ are shorthand notations for
$L_n(\ld_1)\otimes {\bf 1}$ respectively ${\bf 1}\otimes L_n(\ld_2)$.
In section 5, we will specify the
$R$ and $S$ matrix even further, adapting us to the special example
of systems associated with the GD hierarchy. The quantum $L$ matrix
is taken in $End(V\otimes \cal H)$, where $V$ acts as the auxiliary
space (in the teminology of e.g. \cite{TTF}), and $\cal H$ is the
Hilbert space of the quantum system.

The quantum relations (\ref{eq:LLL}) were introduced for the first time
in \cite{Bab}, cf. also \cite{Bab0},
in connection with the quantum Toda theory, and in
\cite{Aleks} for the quantum Wess-Zumino-Novikov-Witten (WZNW) model.
They define what in \cite{AFS} is referred to as lattice
current algebra (LCA), or quantum Kac-Moody algebra on the lattice.
In \cite{NCP}, we introduced them in connection with the
quantization of discrete-time models, namely to quantize mappings of
KdV type.

The compatibility relations of the equations (\ref{eq:RLL},\ref{eq:LSL})
lead to the following set of consistency conditions for $R^{\pm}$ and $S^{\pm}$
\bse \label{eq:RR}
\bea
R^{\pm}_{12}\,R^{\pm}_{13}\,R^{\pm}_{23} &=&
R^{\pm}_{23}\,R^{\pm}_{13}\,R^{\pm}_{12}\  , \label{eq:RRR} \\
R^{\pm}_{23}\,S^{\pm}_{12}\,S^{\pm}_{13} &=&
S^{\pm}_{13}\,S^{\pm}_{12}\,R^{\pm}_{23}\  ,  \label{eq:RSS}
\eea\ese
(i.e. two equations for the + sign, and two equations for the - sign),
where $S^+_{12}=S^-_{21}$. Eq. (\ref{eq:RRR}) is the quantum
Yang-Baxter equation for $R^{\pm}$ coupled with an additional equation
(\ref{eq:RSS}) for $S^{\pm}$. In addition to these relations we need
to impose also
\be \label{eq:RS}
R_{12}^{-}\,S^{-}_{12}\ =\ S^{+}_{12}\,R_{12}^{+}\    ,
\ee
in order to be able to derive suitable commutation relations for the
monodromy matrix of the systems under consideration in this paper.

As we are interested in the canonical structure of discrete-time
integrable systems, i.e. systems for which the time evolution is given
by an iteration of a mapping, we need as explained in \cite{NC}
in addition to (\ref{eq:LLL})
commutation relations involving the discrete-time part of the ZS
system, namely the matrices $M$. These matrices containing quantum
operators, we have a set of non-trivial commutation relations with
the $L$-matrices which are such that
the Yang-Baxter structure is preserved.
Such a proposal, formulated in \cite{NC}, consists
of the following commutation relations in addition to the relations
(\ref{eq:LLL})
\bse \label{eq:MM}\bea
M_{n+1,1} S^+_{12} L_{n,2} &=& L_{n,2} M_{n+1,1}\   , \label{eq:MSL}  \\
L^{\prime}_{n,2} S^-_{12} M_{n,1} &=& M_{n,1} L^{\prime}_{n,2}\   ,
\label{eq:LSM}
\eea \ese
and
\bse\label{eq:MMM} \bea
R^+_{12}\,M_{n,1} M_{n,2} &=& M_{n,2} M_{n,1} R^-_{12} \  , \label{eq:RMM} \\
M^{\prime}_{n,1} S^+_{12} M_{n,2} &=& M_{n,2} M^{\prime}_{n,1}\    .
\label{eq:MSM}
\eea\ese
Some trivial commutation relations need also to be specified, namely
\be \label{eq:MT}
[ M_{n,1},L_{m,2}]=[ M_{n+1,1},L^{\prime}_{m,2}]=[ M_{n,1},M_{m,2}]=
[M^{\prime}_{n+1,1},M_{n,2}]=[M^{\prime}_{n+1,1},L_{n,2}]=0  \ee
($|n-m|\geq 2$), where the brackets denote matrix commutation.
We shall not specify any other commutation relations: they do not
belong to the Yang-Baxter structure, even though they might be
nontrivial, as they depend on the details of specific models. The
relations that we have given are self-contained in the sense that
they are sufficient to show that the commutation relations
(\ref{eq:LLL}) are preserved under the mapping coming from (\ref{eq:ZS}).
Thus the quantum mappings associated with (\ref{eq:ZS}) are canonical
in the sense that they leave the underlying Yang-Baxter structure invariant.
Furthermore, it is shown in \cite{N,NCC} that for specific examples of
quantum mappings that fit into the structure presented here (such as
mappings associated with the lattice KdV and MKdV equations, \cite{PNC}),
there is a unitary operator that generates the mapping,
acting on the quantum phase space of the system.

\paragraph{Quantum Traces}

To obtain quantum invariants of the mappings associated with the
ZS system (\ref{eq:ZS}), we  need to introduce the monodromy matrix
\be
T(\lambda )\ \equiv \stackrel{\longleftarrow}{\prod_{n=1}^P}\
L_n(\lambda )\ .  \label{eq:T}
\ee
The commutation relations for the monodromy matrix are obtained
from the relations for the $L$-matrix, (\ref{eq:LLL}), making use of
the crucial relation (\ref{eq:RS}), and taking into account the
periodic boundary conditions. Thus we obtain
\be
R_{12}^{+}\,T_1 \,S^{+}_{12}\, T_2 \ =\ T_2 \, S^{-}_{12}\,
T_1\, R_{12}^{-}\    .   \label{eq:RTST}
\ee
Eq. (\ref{eq:RTST}) is similar to the commutation relations for
the so-called algebra of currents of a quantum group, cf.
\cite{FRT,RSTS}. Versions of such algebras have been considered
in different contexts, e.g. in connection with boundary conditions
of integrable quantum chains \cite{Skly,Cher,Mezin,KulSkly}.

Following a treatment similar as the one in \cite{Skly}, cf. \cite{Exet},
a commuting parameter-family of operators is obtained by taking
\be
\tau( \lambda )\ =\ tr\left( T(\lambda)K(\lambda )\right)
\  ,     \label{eq:TK}
\ee
for any family of numerical matrices $K(\ld)$ obeying the relations
\be
K_1\,^{t_1\!}\left( (\,^{t_1\!}S^-_{12})^{-1}\right)
\,K_2\,R_{12}^+\ =\ R_{12}^-\,K_2\,^{t_2\!}\left(
(\,^{t_2\!}S^+_{12})^{-1}\right) \,K_1\   . \label{eq:RKSK}
\ee
(We assume throughout that $S_{12}^{\pm}$ and $R_{12}^{\pm}$ are
invertible). The left superscripts $\ ^{t_1}$ and $\ ^{t_2}$ denote
the matrix transpositions with respect to the corresponding factors
1 and 2 in the matricial tensor product.
Expanding (\ref{eq:TK}) in powers of the spectral parameter $\lambda$,
we
obtain a set of commuting observables of the quantum system in terms
of which we can find a common basis of eigenvectors in the associated
Hilbert space.

The quantum mapping for the monodromy matrix, is simply
given by the conjugation
\be
T^{\prime}\,=\,M T M^{-1} \  ,
\label{eq:MTM}
\ee
where $M$ is the $M_n$-matrix at the begin- and end point of the
chain, i.e. $M\equiv M_1=M_{P+1}$. In general, $M$ {\em does not commute}
with the entries of $T$, as in our models $M$ will nontrivially depend on
quantum operators. Thus, the classical invariants obtained by taking the
trace of the monodromy matrix, are no longer invariant on the
quantum level. Therefore, we need to investigate the commutation relations
between the monodromy matrix $T$ and the matrix $M$. These
are given by the equation
\be
(TM^{-1})_1 S_{12}^+ M_2\ =\ M_2 S_{12}^- (TM^{-1})_1   \  ,  \label{eq:TMSM}
\ee
in combination with
\be  \label{eq:MMR}
R^+_{12} M_1 M_2\ =\ M_2 M_1 R^-_{12}\   ,
\ee
where, by abuse of notation, $M_2$ denotes here not the matrix $M$ at
site $n=2$, but the matrix $M_1$ in the second factor of the tensorial product.
(We use the same symbol for both $M$-matrices. It will be
clear from the context which of the $M$-matrices we mean if we use
them below).
It is interesting to note that the set of relations consisting of
(\ref{eq:RTST}) and (\ref{eq:TMSM})
is reminiscent of the relations describing the
cotangent bundle of a quantum group $(T^{\ast}G)_q$, \cite{AF}. However,
in the context of the present paper, we will refer to the algebra ${\cal A}$
generated by $T$ and $M$ with defining relations (\ref{eq:RTST}),
(\ref{eq:TMSM}) and (\ref{eq:MMR}), simply as the quantum mapping algebra.
As a consequence of (\ref{eq:TMSM}), (\ref{eq:MMR}), the commutation relation
(\ref{eq:RTST}) is preserved under the mapping (\ref{eq:MTM}).
Furthermore, eq. (\ref{eq:TMSM}) can be used to find special
solutions for $K$ leading to exact quantum invariants.
Introducing the permutation operator $P_{12}$ acting
in the tensor product of matrices $X(\lambda)$, $Y(\lambda)$, by
\bse \label{eq:P}
\be  P_{12} X_1 Y_2\ =\ X_2 Y_1 P_{12} \    \ ,\   \
(\ld_1=\ld_2)\   , \label{eq:PXY}  \ee
interchanging the factors in the
matricial tensor product in $\Bbb C^N\otimes \Bbb C^N$.
For this operator  we have the trace property
\be  tr_1 P_{12}\,=\,{\bf 1}_2 \  , \label{eq:trP}  \ee \ese
where ${\bf 1}_2$ denotes the identity acting operator on vector space $V_2$.
In fact, introducing a tensor  $K_{12}=P_{12}K_1K_2$,
choosing $\lambda_1 = \lambda_2$, we can take the trace over
both factors in the tensor product, contracting both sides of eq.
(\ref{eq:TMSM}) with $K_{12}$. This leads to an exact quantum invariant
of the form (\ref{eq:TK}), provided that the matrix $K(\ld)$ solves
the condition
\be
tr_1(P_{12}K_2S_{12}^+)={\bf 1}_2 \  ,  \label{eq:PKS}
\ee
in which $tr_1$ denotes the trace over the first factor of the tensor
product. Eq. (\ref{eq:PKS}) is explicitely solved by taking
\be \label{eq:PS}
K_2\,=\,tr_1\left\{ P_{12} \,^{t_1\!}\left( ( \,^{t_1\!}S^+_{12})^{-1}
\right) \right\} \    ,
\ee
and this solution $K(\ld)$ can be shown to obey
also eq. (\ref{eq:RKSK}). Hence, the invariants
obtained from (\ref{eq:TK}) by expanding in powers of the spectral
parameter $\ld$ form a commuting family of operators. For the
quantum mappings considered in \cite{NCP,Exet}, notably mappings
associated with the quantum lattice KdV and modified KdV systems,
we obtain in this way enough invariants to establish integrability.
However, for the more general class of $N\times N$ models presented
in \cite{NC}, we need additional invariants. For this we must
develop the fusion algebra associated with the quantum algebra
given by eqs. (\ref{eq:RTST}), (\ref{eq:TMSM}) and (\ref{eq:MMR}).

\section{Fusion Procedure}
\setcounter{equation}{0}

The structure of the mappings outlined in the previous section holds
for a large class of systems, notably the mappings associated with
the lattice Gel'fand-Dikii hierarchy, \cite{NPCQ}. However, for the
higher members of this hierarchy it is not sufficient to consider
only the trace (\ref{eq:TK}) to generate exact quantum invariants.
In fact, for these systems one needs also higher-order invariants
corresponding roughly to the trace of powers of the monodromy matrix.
The construction of such higher-order commuting families of operators
corresponding to the quantum analogue of the classical
object $tr\left(T^n \right)$, i.e. traces over powers of the monodromy
matrix, is called fusion procedure, cf. \cite{KRS}, and in particular
leads to
the proper definition of quantum determinants and minors, \cite{KS},
cf. also \cite{Drin2,FRT,Skly,KRR}. Some of the results in this
section were also obtained in \cite{Ols} in the special case of
twisted Yangians, cf. also \cite{Nepo}. Fusion is also used to
obtain tensor products of representations of quantum algebras,
notably of affine quantum groups.
This latter connection, however, is not of direct concern to
us here.

\paragraph{Notations}

The objects we need to build for the fusion algebra
are tensorial products of matrices, i.e. they are objects of the form
$A_a=A_{(i_1,i_2,\dots,i_N)}$ and $A_{a,b}=A_{(i_1,\dots,i_n),(j_1,\dots,j_m}$,
depending on a multiple indices $a=(i_1,i_2,\dots,i_n)$ and
$b=(j_1,\dots,j_m)$,
labelling the factors in a tensor product of vector-spaces
$\otimes_{\ar} V_{\ar}$, on which $A_a$ and $A_{a,b}$  act nontrivially.
We can think of these vector spaces as being irreducible modules of the
quantum algebra introduced in the previous section. However, in order to be
concrete, we shall take them to be different copies of $V=\Bbb C^N$,
corresponding to the fundamental representation of specific realisations
of the algebra, that we will consider below.

We can now introduce the following formal scheme of tensorial objects
labelled by multi-indices. First we distinguish between an elementary index,
denoted by $i_1,i_2,\dots$, and multi-indices made up from elementary
indices. The elementary indices correspond to the labels of single vector
spaces, typically irreducible modules of the quantum algebra, whereas the
multi-indices correspond to tensor products of these vector spaces. Next,
we introduce some manipulations on multi-indices allowing us to build objects
that are labelled by its entries. Thus, if $a$ denotes such a multi-index,
i.e. an ordered tuple $a=(i_1,i_2,\dots,i_n)$, then we denote
by $\hat{a}$ the multiple index corresponding to the reverse order of
labels, i.e. $\hat{a}=(i_n,\dots,i_2,i_1)$. Let us denote by $\ell(a)$ the
length of the multi-index $a$, i.e. $\ell(a)=n$ for $a=(i_1,i_2,\dots,i_n)$.
Furthermore, we can join
multi-indices as words in a free algebra on a set, namely if $a$ and $b$
are multi-indices $a=(i_1,i_2,\dots,i_n)$ and $b=(j_1,j_2,\dots,j_m)$,
then we denote by $(ab)$ the multi-index obtained by merging the two
together, i.e. $(ab)=(i_1,i_2,\dots,i_n,j_1,j_2,\dots,j_m)$.

We can now build a hierarchy of tensorial objects, labelled by
multi-indices, starting from an object $A_{ij}$ depending on two
elementary indices. There are two types of objects that are of interest,
namely $one$-muli-index objects $A_a$ and $two$-multi-index objexts
$A_{a,b}$. They are generated from $A_{ij}$ in a recursive way
by following the rules
\be \label{eq:A}
A_{(ab)}=A_b A_{\hat{a},\hat{b}} A_a\    ,
\ee
and
\be \label{eq:AA}
A_{(ab),c}=A_{a,c} A_{b,c}\    \ ,\    \ \ A_{a,(bc)}=A_{a,b} A_{a,c}\    ,
\ee
and adopting the convention that we take $A_i\equiv {\bf 1}_i$, i.e. the
objects $A$ depending on a single
elementary index act as the unit matrix on the corresponding vector space.
(Note also that one has to distinguish the two-multi-index object $A_{a,b}$
from the one-multi-index object $A_{(ab)}$, although for elementary indices
we take $A_{(ij)}=A_{i,j}$).
Thus for example, we in building these tensorial objects, we will obtain
$$ A_{i_1,\dots,i_n}\,=\,\left( A_{i_{n-1},i_n}\right)
\left( A_{i_{n-2},i_n} A_{i_{n-2},i_{n-1}}\right) \cdots
\left( A_{i_1,i_n} A_{i_1,i_{n-1}} \cdots A_{i_1,i_2}\right)\   , $$
and
$$ A_{(i_1,\dots,i_n),(j_1,\dots,j_m)}\,=\,
\left( A_{i_1,j_1} \cdots A_{i_n,j_1}\right) \cdots
\left( A_{i_1,j_m} \cdots A_{i_n,j_m}\right) \    .
$$
However, the recursive relations (\ref{eq:A}) and (\ref{eq:AA}) will be more
useful for us than these explicit expressions.

\paragraph{Fundamental Relations}

Let us take for $A$ now objects like $R^{\pm}_{12}$, obeying the Yang-Baxter
equation (\ref{eq:RRR}) or $S^{\pm}_{12}$ obeying eq. (\ref{eq:RSS}), as well
as the important relation (\ref{eq:RS}). Then, we
can generate, using the prescriptions (\ref{eq:A}), multi-index objects
$R_a$, $S_a$, depending on a single multi-index, or objects $R_{a,b}$
and $S_{a,b}$ depending on a double multi-index according to (\ref{eq:AA}).
For these objects we can derive the Yang-Baxter type of relations
\bse \label{eq:rr}
\bea
R^{\pm}_{a,b}R^{\pm}_{a,\hat{c}}R^{\pm}_{\hat{b},\hat{c}} &=&
R^{\pm}_{\hat{b},\hat{c}}R^{\pm}_{a,\hat{c}}R^{\pm}_{a,b} \   , \label{eq:rrr}
\\
R^{\pm}_{b,c}S^{\pm}_{a,b}S^{\pm}_{a,\hat{c}} &=&
S^{\pm}_{a,\hat{c}}S^{\pm}_{a,b}R^{\pm}_{b,c} \   .
\label{eq:rss}
\eea
\ese
In addition, we can derive the identities
\bse \label{eq:Rr}
\bea
R^{\pm}_aR^{\pm}_{a,b} &=& R^{\pm}_{\hat{a},b}R^{\pm}_a\   , \label{eq:rra} \\
R^{\pm}_aR^{\pm}_{b,\hat{a}} &=& R^{\pm}_{b,a}R^{\pm}_a\   , \label{eq:rrb} \\
R^{\pm}_aS^{\pm}_{b,a} &=& S^{\pm}_{b,\hat{a}}R^{\pm}_a\   . \label{eq:rrc}
\eea
\ese
In other words, $R^{\pm}_a$ reverses the multi-index $a$ when it is pulled
through an $R^{\pm}_{a,b}$ or an $S^{\pm}_{a,b}$.
Then, there are relations that can be derived starting from (\ref{eq:RS}),
namely
\be \label{eq:SR}
R^{-}_{\hat{a},b}S^{+}_{(ba)}\,=\,S^{+}_{(ab)}R^{+}_{\hat{a},b}\   ,
\ee
as well as
\be \label{eq:sr}
S^{+}_aR^{+}_a\,=\,R^{-}_aS^{+}_{\hat{a}}\   .
\ee
In other words, $S^{+}_a$ reverses the sign when it is pulled
through an $R^{\pm}_a$ or an $R^{\pm}_{a,b}$.
Eqs. (\ref{eq:rr})-(\ref{eq:sr}), which are proven in Appendix A
provide us with the complete set of
relations that we need to be able to define the fusion algebra for the
quantum mappings.

\paragraph{Fusion algebra}

Apart from the multi-index tensor objects that can be generated from
an elementary 2-index object, like $S^{\pm}_{12}$ or $R^{\pm}_{12}$,
we have multi-index objects that are generated from a single-index
object like $T_1$ or $M_1$. These are iteratively defined
by the relations
\be  \label{eq:tt}
T_{(ab)}\,=\,R^+_{\hat{a},b} T_a S^+_{a,\hat{b}} T_b\   \ ,\   \
M_{(ab)}\,=\,M_a M_b\     .
\ee
{}From the relations (\ref{eq:RTST}),(\ref{eq:RMM}),(\ref{eq:TMSM}) for
objects $T$ and $M$, depending on an elementary index, one can
derive by iteration the following set of relations
\bse  \label{eq:rtt}
\bea
R^+_{\hat{a},b} T_a S^+_{a,\hat{b}} T_b &=& T_b S^-_{\hat{a},b} T_a
R^-_{a,\hat{b}}\   ,
\label{eq:rtst} \\
R^+_{a,b} M_a M_{\hat{b}} &=& M_{\hat{b}} M_a R^-_{a,b}\   , \label{eq:rmm} \\
\left[ T_a M_a^{-1} (S^+_{\hat{a}})^{-1}\right] S^+_{a,b} M_b &=&
M_b S^-_{\hat{a},b} \left[ T_a M_a^{-1} (S^+_{\hat{a}})^{-1}\right]\   ,
\label{eq:tmsm}    \\
R^+_a M_a &=& M_{\hat{a}} R^-_a\    . \label{eq:rm}
\eea
\ese
Furthermore, as a consequence of (\ref{eq:tt}), with the use of (\ref{eq:Rr}),
we have that $T_a$ is of the form
\be
T_a\, =\, R^+_a T^0_a\,=\,T^0_{\hat{a}} R^-_a \    \ ,\   \
T^0_{(ab)}\,=\,T^0_a S^+_{a,b} T^0_b\   . \label{eq:rt}
\ee
The quantum mapping obtained by iteration of (\ref{eq:MTM}), with the use
of (\ref{eq:tt}), for the multi-index monodromy matrix $T_a$, is given by
\be \label{eq:tm}
T^{\prime}_a S^+_{\hat{a}} M_a\,=\,M_{\hat{a}} S^+_a T_a\   .
\ee
Eqs. (\ref{eq:rtt}), together with the definitions (\ref{eq:tt}), and the
mapping (\ref{eq:tm}),
define a complete set of relations by which we can described higher-order
tensor products of the algebra generated by $M$ and $T$ as given by the
relations (\ref{eq:RTST})-(\ref{eq:MMR}). We shall refer to eqs.
(\ref{eq:rtt}) as the defining relations for the fusion algebra for the
quantum mappings described by (\ref{eq:MTM}). After giving a proof of these
equations, we will use them in order
to generate higher-order commuting families of exact quantum invariants of the
mappings defined by (\ref{eq:tm}).

\paragraph{Proof of relations (\ref{eq:rtt})} To prove eqs. (\ref{eq:rtt})
it is most convenient to break down the relations for multi-index
objects depending on joint indices $(ab)$  into separate parts,
assuming that the relations hold for these parts (by induction). Thus,
to prove e.g. eq. (\ref{eq:rtst}) we can perform the following sequence
of steps
\bea
&& R^+_{\widehat{(ab)},c} T_{(ab)} S^+_{(ab),\hat{c}} T_c\ =    \nn \\
&&\ \ = R^+_{\hat{b},c} R^+_{\hat{a},c} R^+_{\hat{a},b} T_a S^+_{a,\hat{b}}
T_b S^+_{a,\hat{c}}  S^+_{b,\hat{c}} T_c  \nn \\
&&\ \ = R^+_{\hat{b},c} R^+_{\hat{a},c} T_b S^-_{\hat{a},b} T_a
R^-_{a,\hat{b}} S^+_{a,\hat{c}} S^+_{b,\hat{c}} T_c  \nn \\
&&\ \ = R^+_{\hat{b},c} R^+_{\hat{a},c} T_b S^-_{\hat{a},b} T_a
S^+_{b,\hat{c}} S^+_{a,\hat{c}} T_c R^-_{a,\hat{b}}   \nn \\
&&\ \ = R^+_{\hat{b},c} T_b S^+_{b,\hat{c}} S^-_{\hat{a},b}
R^+_{\hat{a},c} T_a S^+_{a,\hat{c}} T_c R^-_{a,\hat{b}}   \nn \\
&&\ \ = R^+_{\hat{b},c} T_b S^+_{b,\hat{c}} T_c S^-_{\hat{a},b}
S^-_{\hat{a},c} T_a R^-_{a,\hat{c}} R^-_{a,\hat{b}}   \nn \\
&&\ \ = T_c S^-_{\hat{b},c} T_b S^-_{\hat{a},c} S^-_{\hat{a},b}
R^-_{b,\hat{c}} T_a R^-_{a,\hat{c}} R^-_{a,\hat{b}}   \nn \\
&&\ \ = T_c S^-_{(\hat{b}\hat{a}),c} T_b S^-_{\hat{a},b} T_a
R^-_{a,\hat{b}} R^-_{a,\hat{c}} R^-_{b,\hat{c}}   \nn \\
&&\ \ = T_c S^-_{\widehat{(ab)},c} T_{(ab)} R^-_{(ab),\hat{c}}\   .
\label{eq:nn1}
\eea
In a similar fashion one can show that
\be
R^+_{\hat{a},(bc)} T_a S^+_{a,\widehat{(bc)}} T_{(bc)}\ =\
T_{(bc)} S^-_{\hat{a},(bc)} T_a R^-_{a,\widehat{(bc)}}\   ,
\label{eq:nn2}  \ee
and eq. (\ref{eq:rtst}) follows from (\ref{eq:nn1}) and (\ref{eq:nn2}) by
induction. The consistency of the mapping (\ref{eq:tm}) with the relations
(\ref{eq:rtt}) is checked by the following calculation
\bea
T^{\prime}_{(ab)} &=&
R^+_{\hat{a},b} T^{\prime}_a S^+_{a,\hat{b}} T^{\prime}_b\ =  \nn \\
&=& R^+_{\hat{a},b} M_{\hat{a}} S^+_a \left[ T_a M_a^{-1} (S^+_{\hat{a}})^{-1}
S^+_{a,\hat{b}} M_{\hat{b}}\right] S^+_b T_b M_b^{-1} (S^+_{\hat{b}})^{-1}
 \nn \\
&=& R^+_{\hat{a},b} M_{\hat{a}} M_{\hat{b}} S^+_a S^-_{\hat{a},\hat{b}}
 S^+_b \left[ T_a M_a^{-1} (S^+_{\hat{a}})^{-1}\right]
\left[ T_b M_b^{-1} (S^+_{\hat{b}})^{-1}\right]  \nn \\
&=& M_{\hat{b}} M_{\hat{a}} S^+_{(ab)} R^+_{\hat{a},b} T_a S^-_{\hat{b},a}
\left[ T_b M_b^{-1} (S^+_{\hat{b}})^{-1}\right] M_a^{-1}
(S^+_{b,a})^{-1} (S^+_{\hat{a}})^{-1}   \nn \\
&=& M_{\widehat{(ab)}} S^+_{(ab)} T_{(ab)} M_{(ab)}^{-1}
(S^+_{\widehat{(ab)}})^{-1}\    . \label{eq:nn7}
\eea
Finally, to prove (\ref{eq:tmsm}), we note that
\be R^+_a \left( T\,M^{-1}\right)_a  S^+_{a,b} M_b\,=\,
M_b S^-_{\hat{a},b} R^+_a \left( T\,M^{-1}\right)_a \  ,
\label{eq:nn3} \ee
in which we use the following notation for the
iterate of the single-index matrix $TM^{-1}$
\[  \left( T\,M^{-1}\right)_{(ab)} \,=\,
\left( T\,M^{-1}\right)_a \left( T\,M^{-1}\right)_b\   ,  \]
together with the relation
\be
R^+_{a} \left( T\,M^{-1}\right)_{a}\,=\,
T_a M_a^{-1} (S^+_{\hat{a}})^{-1}\    . \label{eq:nn4} \ee
This last relation can be proven inductively by the folowing
steps
\bea
R^+_{(ab)} \left( T\,M^{-1}\right)_{(ab)} &=&
R^+_{\hat{a},b} R^+_a R^+_b \left( T\,M^{-1}\right)_a
\left( T\,M^{-1}\right)_b   \nn \\
&=& R^+_{\hat{a},b} T_a M_a^{-1} (S^+_{\hat{a}})^{-1}
T_b M_b^{-1} (S^+_{\hat{b}})^{-1}   \nn \\
&=& R^+_{\hat{a},b} T_a S^+_{a,\hat{b}} T_b M_b^{-1} M_a^{-1}
(S^+_{\hat{b}})^{-1} (S^+_{b,a})^{-1} (S^+_{\hat{a}})^{-1}
\,=\,T_{(ab)} M_{(ab)}^{-1} (S^+_{\widehat{(ab)}})^{-1}\   ,   \label{eq:nn5}
\eea
when we break up the indices into smaller parts.  This concludes the
induction argument in the proof of eqs. (\ref{eq:rtt}).

\section{Quantum Invariants}
\setcounter{equation}{0}

\newcommand{\TT}{{\cal T}}
\newcommand{\SS}{{\cal S}}
\newcommand{\MM}{{\cal M}}
\newcommand{\RR}{{\cal R}}
\newcommand{\KK}{{\cal K}}
\newcommand{\PP}{{\cal P}}
\newcommand{\LL}{{\cal L}}

In order to make the formulas derived in the previous section more transparent
we introduce for convenience some new objects, which seem to be slightly
more natural. Thus, we define
\bea
\TT_a &\equiv& S^+_a T_a\   \ ,\   \
\MM_a \equiv S^+_{\hat{a}} M_a\    ,  \nn \\
\SS^+_{a,b} &\equiv& S^+_{\hat{b}} S^+_{a,b} (S^+_{\hat{b}})^{-1}
\    \ , \    \  \SS^-_{a,b} \equiv \SS^+_{b,a}\   ,
 \label{eq:def} \\
\RR^+_{a,b} &\equiv& S^+_{\hat{a}} S^+_b R^+_{a,b} (S^+_{\hat{a}})^{-1}
(S^+_b)^{-1}\   \ , \    \  \RR^-_{a,b}\equiv R^-_{a,b}\   .  \nn
\eea
With these notations it is easily verified that eqs. (\ref{eq:rr})
still hold, but now in terms of $\RR^{\pm}_{a,b}$ and
$\SS^{\pm}_{a,b}$ instead of the original $R^{\pm}_{a,b}$ resp.
$S^{\pm}_{a,b}$. However, eq. (\ref{eq:SR}) now reduces to
\be \RR^-_{a,\hat{b}} \SS^-_{a,b}\,=\,
\SS^+_{a,b} \RR^+_{a,\hat{b}}\   ,  \label{eq:crs}  \ee
for which combination we have a `twisted' Yang-Baxter relation
\be
\RR^-_{a,\hat{b}} \left( \SS^+_{a,c} \RR^+_{a,\hat{c}}\right)
\SS^-_{a,b} \left( \SS^+_{b,c} \RR^+_{b,\hat{c}}\right)\ =\
\left( \SS^+_{b,c} \RR^+_{b,\hat{c}}\right) \SS^+_{a,b}
\left( \SS^+_{a,c} \RR^+_{a,\hat{c}}\right)
\RR^+_{a,\hat{b}}\    .  \label{eq:twYB}
\ee
Furthermore, we have from eqs. (\ref{eq:Rr}) and (\ref{eq:sr})
\bse \label{eq:cRr} \bea
\RR^{\pm}_{\hat{a},b} R^-_a &=& R^-_a \RR^{\pm}_{a,b}\   ,
\label{eq:crra} \\
\RR^{\pm}_{a,b} R^-_b &=& R^-_b \RR^{\pm}_{a,\hat{b}}\   ,
\label{eq:crrb} \\
\SS^{\pm}_{b,\hat{a}} R^-_a &=& R^-_a\,\SS^{\pm}_{b,a}\   .
\label{eq:csr} \eea \ese
Using (\ref{eq:def}) we readily obtain from eqs. (\ref{eq:rtt})
\bse \label{eq:crtt}
\bea
\RR^+_{\hat{a},b} \TT_a \SS^+_{a,\hat{b}} \TT_b &=&
\TT_b \SS^-_{\hat{a},b} \TT_a \RR^-_{a,\hat{b}}\   ,
\label{eq:crtst} \\
\RR^+_{a,b} \MM_a \MM_{\hat{b}} &=& \MM_{\hat{b}} \MM_a \RR^-_{a,b}\   ,
\label{eq:crmm} \\
\TT_a \MM_a^{-1} \SS^+_{a,b} \MM_b &=&
\MM_b \SS^-_{\hat{a},b} \TT_a \MM_a^{-1} \    .
\label{eq:ctmsm}
\eea
\ese
The mapping (\ref{eq:tm}) adopts the more natural form
\be \label{eq:ctm}
\TT_a^{\prime} \MM_a\,=\,\MM_{\hat{a}} \TT_a\   .
\ee
It is interesting to note the near resemblance between eqs. (\ref{eq:crtt})
and the original equations (\ref{eq:RTST})-(\ref{eq:MMR}).  Hence,
in this form, the fusion algebra, defined by these relations, has the
most convenient form to calculate quantum invariants following a
prescription analogous to the construction of eqs. (\ref{eq:TK}) and
(\ref{eq:PS}), together with (\ref{eq:RKSK}).
We also note the relation
\be
\TT_{(ab)}\,=\,S^+_{(ab)} T_{(ab)}\,=\,\left( \SS^+_{\hat{a},\hat{b}}
\RR^+_{\hat{a},b}\right) \TT_a \SS^+_{a,\hat{b}} \TT_b\    .
\ee

\paragraph{Projectors}

The relations (\ref{eq:cRr}) and (\ref{eq:crtt}) hold for arbitrary choice
of the values of the spectral parameters $\ld_a=(\ld_1,\dots,\ld_n)$,
$\ld_b=(\mu_1,\dots,\mu_m)$, associated
with the factors denoted by the multi-indices $a=(i_1,\dots,i_n)$ and
$b=(j_1,\dots,j_m)$ in the tensor products. In order  to construct
quantum invariants, we need now to impose certain relations between
the values of the different $\ld_1,\dots,\ld_n$, and between the
$\mu_1,\dots,\mu_m$.
We now make the crucial assumption that for special choices of the
spectral parameters $\ld_a=(\ld_1,\dots,\ld_n)$ in
$R^{\pm}_a$ for $a=(i_1,\dots,i_n)$,
the matrix $R^-_a$ becomes a projector,
\be \label{eq:proj}  (R^-_a)^2\,=\,R^-_a\  . \ee
This condition is satisfied in particular for models obeying the
so-called ``regularity condition'', \cite{TTF}. For example, in the
GD class of models that we consider in section 5, we have
\be R^-_{12}(\ld,\ld \pm h)\,=\,{\bf 1} \pm P_{12}\   , \label{eq:rpp} \ee
for some value $h\in \Bbb C$.
In that case, cf. also \cite{KS},  we obtain from $R^-_{(1,\dots,n)}$
a realization of the
completely (anti)symmetric tensor
\be
 P^{\pm}_{(1,\dots,n)}\,=\,\frac{1}{n!}\sum_{\sigma\in S^n}\,(\pm 1)^{\sigma}
P_{\sigma}\   ,  \label{eq:anti}  \ee
$P_{\sigma}$ denoting the represesentation of the symmetric group
$S^n$ on the $n$-fold tensor product $(\Bbb C^N)^{\otimes n}$. In a more
general context, we can think of these projectors as projecting out
the various irreducible blocks in the tensor product of modules on the
quantum algebra.

Furthermore, we note that from (\ref{eq:rt}) we have
\be \TT_a\,=R^-_a\left( S^+_{\hat{a}}T^0_a\right)\,
=\,\left( S_a^+T^0_{\hat{a}}\right) R_a^-\   \ ,\   \ R^-_a\MM_a\,=\,
\MM_{\hat{a}} R^-_a\   ,  \label{eq:crt}
\ee
using also eq. (\ref{eq:sr}). It is suggestive in the case of the special
choice of parameters for
which we have eq. (\ref{eq:proj}) to  refer to the  objects
$\TT_a=R^-_a\TT_aR^-_a$ simply as {\em quantum minors}. In particular for
situations in which we have (\ref{eq:rpp}), i.e. when  $R^-_a$ projects out an
$(n+1-\ell(a))$-dimensional subspace of the tensor product
$V^{\otimes n}$ of the auxiliary vector space $V=\Bbb C^N$,
the coefficients of $\TT_a$ become the actual quantum minors.

\paragraph{Construction of Higher-order Invariants}

We now proceed by deriving the higher-order quantum invariants  of the
mapping, which is now encoded in eq. ({\ref{eq:ctm}). These are
obtained by making direct use of eq. (\ref{eq:ctmsm}).

In the spirit of the above construction of multi-index objects, one can
introduce multi-index permutation operators $\PP_{a,b}$, generated
from the elementary object $P_{12}$, (the permutation matrix in the
matricial
tensor product $\Bbb C^N\otimes \Bbb C^N$) following the prescription
in section 3.  These operators $\PP_{a,b}$ have the following
properties when acting on arbitrary multi-index objects
$X_a=X_a(\ld_a)$,
\bea
tr_a \PP_{a,b}\,=\,tr_a \PP_{b,a}\,=\,{\bf 1}_b\    \ &,&\    \
\PP_{a,b} X_a\,=\,X_{\hat{b}} \PP_{a,b}\    \ ,\    \
\PP_{a,b} X_b\,=\,X_{\hat{a}} \PP_{a,b}\    \ ,\    \ \PP_{a,b}
\PP_{\hat{b},\hat{a}}={\bf 1}_{a,b}\    , \label{eq:pp}  \\
&&( \ell(a)=\ell(b)\ \ ,\ \ \ld_a=\ld_{\hat{b}} )\   ,  \nn
\eea
in which $\ld_a\equiv (\ld_1,\dots , \ld_n)$ denotes the collection of
spectral parameters on which $X_a$ depends, and denoting by $tr_a$ the multiple
trace over all factors in the tensor product labelled by the multi-index $a=
(i_1,\dots,i_n)$.

We have now the following statement
\paragraph{Proposition 1:} If $\ld_a$ is chosen such that $R^-_a$
is a projection matrix on the $\ell(a)$-fold tensor product of
vector spaces $V_i$, ($i=1,\dots,n$), the the family of operators
\be \tau^{(n)}(\ld_a) \equiv tr_a(\KK_a\TT_a) \label{eq:ctau} \ee
are exact invariants of the mapping (\ref{eq:ctm}), provided that
$\KK_a$ solves the equation
\be \label{eq:trPKS}
tr_a\left( \PP_{a,\hat{b}} \KK_b \SS^+_{a,\hat{b}}\right)\,=\,{\bf 1}_b\   ,
\ee
where $\ell(a)=\ell(b)=n$, $\ld_a=\ld_b$.\\
\\
{\em Proof:} Contracting both sides of eq. (\ref{eq:ctmsm}) with
$\PP_{a,\hat{b}}\KK_a\LL_b$, in which $\KK_a$ and $\LL_b$ are
(numerical) matricial tensors that need to be
determined, and using (\ref{eq:pp}), we find the equality
\bea
tr_{a,b} \left\{ \PP_{a,\hat{b}} \KK_a\LL_b \TT_a\MM_a^{-1} \SS^+_{a,b}\MM_b
\right\}
&=& tr_b \left\{ \KK_b\TT_b\MM_b^{-1} tr_a\left( \PP_{a,\hat{b}}\LL_b
\SS^+_{a,b}\right) \MM_b \right\}\,=    \nn \\
=\,tr_{a,b} \left\{ \PP_{a,\hat{b}} \KK_a\LL_b \MM_b
\SS^-_{\hat{a},b}\MM_a^{-1}
\TT_a^{\prime} \right\}
&=& tr_a \left\{ \LL_a\MM_a tr_b\left( \PP_{a,\hat{b}}\KK_a
\SS^-_{\hat{a},b}\right) \MM_{\hat{a}}^{-1}\TT_a^{\prime}  \right\}\  ,    \nn
\eea
where $tr_{a,b}=tr_atr_b$.
Using the fact that $R_a^-$ is a projector, and noting eqs.
({\ref{eq:rt}), it is easily seen that we can choose the conditions
\bse \label{eq:trPLS} \bea
tr_a\left( \PP_{a,\hat{b}} \LL_b \SS^+_{a,b}\right) &=& R^-_b\   ,   \\
tr_b\left( \PP_{a,\hat{b}} \KK_a \SS^-_{\hat{a},b}\right) &=& {\bf 1}_a\
  .   \eea  \ese
Both equations can be solved simultaneously if
\[ \LL_a\,=\,\KK_a R^-_a\   ,   \]
using the relation  (\ref{eq:csr}).
In that case, the above equality reduces to
\[ tr_a\left( \KK_a\TT_a\MM_a^{-1} R^-_a \MM_a\right)\,=\,
tr_a\left( \KK_a R^-_a \MM_a \MM_{\hat{a}}^{-1} \TT_a^{\prime} \right)\   . \]
Using again (\ref{eq:rt}), we now note that
\[
R_a\MM_a\MM_{\hat{a}}^{-1}R_a^-
\,=\,\MM_{\hat{a}} \left( R_a^-\right)^2 \MM_a^{-1}\,=\,
\MM_{\hat{a}} R_a^- \MM_a^{-1}\,=\,
R_a^-\MM_a\MM_a^{-1}\,=\,R_a^-\  ,  \]
and similarly
\[  R_a\MM_a^{-1} R_a^-\MM_a \,=\,
\MM_{\hat{a}}^{-1} \left( R_a^-\right)^2 \MM_a\,=\,
\MM_{\hat{a}}^{-1} R_a^- \MM_a\,=\,
R_a^-\MM_a^{-1}\MM_a\,=\,R_a^-\  ,  \]
making use of (\ref{eq:crt}) and the fact that $R_a^-$ is a
projector, leading to
\be  tr_a\left( \KK_a\TT_a\right)\,=\,tr_a\left(
\KK_a\TT_a^{\prime}\right) \   . \label{eq:trinv}
\ee
\hfill \rule{2mm}{3mm} \\

\paragraph{Remark:} We note that the requirement for invariance of
operators of the form (\ref{eq:ctau}) in the case that the $R^-_a$
are {\em not} projectors can still be met by taking
\bse  \bea
tr_a\left( \PP_{a,\hat{b}} \LL_b \SS^+_{a,b}\right) &=& {\bf 1}_b\   ,  \nn  \\
tr_b\left( \PP_{a,\hat{b}} \KK_a \SS^-_{\hat{a},b}\right) &=&
R^-_{\hat{a}}\  , \nn   \eea  \ese
instead of (\ref{eq:trPLS}). Both relations can be simultaneously
solved by taking now
\[ \KK_a\,=\,\LL_a R^-_{\hat{a}}\   . \]
However, the resulting invariants will
typically factorize into lower order invariants, because of the
combination $R^-_{\hat{a}} R^-_a$ that will occur when contracting
with $T_a$, in combination with the unitarity condition
\[ R_{12}(\ld_1,\ld_2) R_{21}(\ld_2,\ld_1)\,=\,{\bf 1}\  ,
\]
that is applicable to many of the well-known quantum models, in
particular the ones considered in section 5. \\

It now remains to be shown that the higher-order trace objects $\tau^{(n)}$
defined above yield a commuting family of quantum operators. This is
ensured by the following statement
\paragraph{Proposition 2:} In order for (\ref{eq:ctau}) to yield a
commuting family of operators it is sufficient that the following
relation holds
\be \label{eq:crksk}
\KK_a\,^{t_a\!}\left( (\,^{t_a\!}\SS^-_{\hat{a},b})^{-1}\right) \KK_b
\RR^+_{\hat{a},b}\,=\, \RR^-_{a,\hat{b}} \KK_b\,^{t_b\!}\left(
(\,^{t_b\!}\SS^+_{a,\hat{b}})^{-1}\right) \KK_a\    ,
\ee
(the left-superscripts $t_a,t_b$ denote matrix transposition in the
factors corresponding  to the labels $a,b$ in the tensor products).
If, in addition, the spectral parameters $\ld_a$ and $\ld_b$ are
chosen such that
$R^-_a$ and $R^-_b$ are projectors, (\ref{eq:proj}), then it is
sufficient that (\ref{eq:crksk}) holds  modulo a multiplication from
the left and the right by factors $R^-_a$ and $R^-_b$. \\
{\em Proof:}
In order to derive eq. (\ref{eq:crksk}), let us give an argument similar
to the one given by Sklyanin in \cite{Skly}, cf. also \cite{Exet}.
Denoting by $\tau_a$,$\tau_b$ the invariants (\ref{eq:ctau}) evaluated
at different
values $\ld_a$ resp. $\ld_b$ of the spectral parameters, we have
\bea
\tau_a \, \tau_b &=& tr_a\left( \TT_a\KK_a\right) \,
tr_b\left( \TT_b\KK_b\right)\
=\ tr_{a,b}\left\{ \TT_a\KK_a\, \,^{t_b\!}\TT_b\,^{t_b\!}\KK_b\right\} \nn \\
&=& tr_{a,b}\left\{ \,^{t_b\!}(\TT_a\, \SS^+_{a,\hat{b}}\TT_b)\ ^{t_a\!}(
\,^{t_a\!}\KK_a\, \,^{t_{a}\!}\left( (\,^{t_b\!}\SS^+_{a,\hat{b}})^{-1}\right)
\,^{t_b\!}\KK_b)\,\right\} \nn \\
&=& tr_{a,b}\left\{ \left( \RR^+_{\hat{a},b}\right)^{-1}\,\TT_b\,
\SS^-_{\hat{a},b}\TT_a\,\RR_{a,\hat{b}}^-\
^{t_{ab}\!}\left[ \,^{t_a\!}\KK_a\, \,^{t_{a}\!}\left(
(\,^{t_b\!}\SS^+_{a,\hat{b}})^{-1}\right) \,^{t_b\!}\KK_b \right]\,\right\}\
\label{eq:D1}\\
&=& tr_{a,b}\left\{ \,\TT_b\, \SS^-_{\hat{a},b}\TT_a\ ^{t_{ab}\!}\left[
\left( \,^{t_{ab}\!}\RR^+_{\hat{a},b}\right)^{-1}
\,^{t_a\!}\KK_a\, \,^{t_{a}\!}\left( (\,^{t_b\!}\SS^+_{a,\hat{b}})^{-1}\right)
\,^{t_b\!}\KK_b \,^{t_{ab}\!}\RR_{a,\hat{b}}^- \right] \right\}\  ,
\nn \eea
whereas on the other hand we have
\bea
\tau_b \, \tau_a &=& tr_b\left( \TT_b\KK_b\right) \,
tr_a\left( \TT_a\KK_a\right) \nn \\
&=& tr_{a,b}\left\{ \,^{t_a\!}\left( \TT_b\,
\SS^-_{\hat{a},b}\TT_a\right)\ ^{t_b\!}\left[ \,
^{t_b\!}\KK_b\, \,^{t_{b}\!}\left( (\,^{t_a\!}\SS^-_{\hat{a},b})^{-1}\right)
\,^{t_a\!}\KK_a\right] \right\}
\label{eq:D2} \\
&=& tr_{a,b}\left\{ \TT_b\, \SS^-_{\hat{a},b}\TT_a\ ^{t_{ab}\!}(
\,^{t_b\!}\KK_b\, \,^{t_{b}\!}\left( (\,^{t_a\!}\SS^-_{\hat{a},b})^{-1}\right)
\,^{t_a\!}\KK_a)\,\right\} \    , \nn
\eea
which leads to eq. (\ref{eq:crksk}) identifying (\ref{eq:D1}) and
(\ref{eq:D2}). Clearly, in the case that $R^-_a$ and $R^-_b$ are
projectors, taking note of (\ref{eq:crt}), it is sufficient that a
weaker condition holds, namely (\ref{eq:crksk}) multiplied from the left
and the right by these projectors.  \hfill \rule{2mm}{3mm} \\

\newcommand{\pp}{^{\prime}}

It now remains to be established that the solution of (\ref{eq:trPKS}),
which leads to exact quantum invariants of the mapping, also obeys
(\ref{eq:crksk}). This is stated by the following
\paragraph{Proposition 3:} If $\ld_a$ is chosen such that $R^-_a$ is
a projector, then the family of operators (\ref{eq:ctau}),
where $\KK_a$ is a solution of (\ref{eq:trPKS}), forms a commuting
family of quantum operators, i.e.
\be \label{eq:tautau}
\left[ \tau^{(n)}(\ld_a)\,,\,\tau^{(m)}(\mu_b)\right] \,=\,0
\    ,
\ee
for $n=\ell(a)$, $m=\ell(b)$.\\
\\
{\em Proof:}
The solution of eq. (\ref{eq:trPKS}) is given by
\footnote{In fact, for tensor objects of the
form $X_{a,b}$,$\ Y_{a,b}$,
one can introduce an associative twisted product, (for $\ell(a)=\ell(b)$),
\[ X_{a,b} \ast Y_{a,b} \equiv \,^{t_b\!}\left( \,^{t_b\!}X_{a,b}
\,^{t_b\!}Y_{a,b}\right) \,=\,
\,^{t_a\!}\left( \,^{t_a\!}Y_{a,b}
\,^{t_a\!}X_{a,b}\right) \  . \]
Then, an inverse with respect to the product $\ast$ is given by
\[ X_{a,b}^{-1_{\ast}}\,=\,\,^{t_b\!}\left(
(\,^{t_b\!}X_{a,b})^{-1} \right) \,=\,
\,^{t_a\!}\left( (\,^{t_a\!}X_{a,b})^{-1} \right) \  .  \]
The solution $\KK_a$ of (\ref{eq:trPS}) is the
contraction of the $\ast$-inverse of $\SS^+_{a\pp,\hat{a}}$. }
\be
\KK_a\,=\,tr_{a\pp}\left\{ \PP_{a\pp,\hat{a}}\,^{t_a\!}\left(
(\,^{t_a\!}\SS^+_{a\pp,\hat{a}})^{-1}\right) \right\} \   .  \label{eq:trPS}
\ee
This is easily verified by the  observation that
\bea
tr_b &&\left\{ \PP_{b,\hat{a}} \KK_a \SS^+_{b,\hat{a}}\right\}\,=\,
tr_{a\pp,b}\left\{ \PP_{b,\hat{a}} \PP_{a\pp,\hat{a}}
( \SS^+_{a\pp,\hat{a}})^{-1_{\ast}} \SS^+_{b,\hat{a}}\right\}  \nn \\
&&=\,tr_{a\pp,b}\left\{ \PP_{a\pp,\hat{a}} \PP_{b,\hat{a\pp}}
( \SS^+_{a\pp,\hat{a}})^{-1_{\ast}} \SS^+_{b,\hat{a}}\right\}
\,=\,tr_{a\pp}\left\{ \PP_{a\pp,\hat{a}}\,
 tr_b \left( \PP_{b,\hat{a\pp}}
( \SS^+_{a\pp,\hat{a}})^{-1_{\ast}} \SS^+_{b,\hat{a}}\right) \right\}
 \nn \\
&&=\,tr_{a\pp}\left\{ \PP_{a\pp,\hat{a}}\, \left(
\SS^+_{a\pp,\hat{a}} \ast ( \SS^+_{a\pp,\hat{a}} )^{-1_{\ast}} \right)
\right\} \,=\,tr_{a\pp}\left( \PP_{a\pp,\hat{a}} {\bf 1}_{a\pp,\hat{a}}
\right) \,=\,{\bf 1}_a\   ,  \nn
\eea
In order to have a commuting family of invariants, it is sufficient according
to proposition 2 that $\KK_a$ obeys eq. (\ref{eq:crksk}) projected by
factors $R^-_a$ and $R^-_b$. This condition is verified by the solution
(\ref{eq:trPS}) for the given choice of spectral parameters.
This is checked by
simply inserting $\KK_a$ into (\ref{eq:crksk}) and by using the
relations (\ref{eq:crtt}) together with (\ref{eq:crs}) to verify that
the equation is satisfied. Details of this calculation are provided
in Appendix B.
\hfill \rule{2mm}{3mm} \\

As a corollory of proposition 3, we recover under special
circumstances the construction of quantum determinants associated
with the algebra ${\cal A}$, namely when $R^-_a$ is a fully
antisymmetric tensor projecting out a one-dimensional
subspace in the tensor product of $\Bbb C^N$. In that case, we will call
the length $\ell(a)$ of the corresponding multi-index $a$ {\em maximal},
and the corresponding minors $\TT_a$ of maximal length will be referred to
as the quantum determinants of the model. We will come back to this in
section 5. \\

Thus, we have now a complete and general construction of
commutative families of
exact quantum invariants of mappings (\ref{eq:ctm})
associated with the Yang-Baxter structure (\ref{eq:crtt}).
We note that the requirement of having not only a commuting family
of operators (for which any solution of eq. (\ref{eq:crksk}) suffices),
but in addition for these operators to be invariants under the mapping,
is strong enough to uniquely determine the invariant family, i.e. it
forces us to consider the specific solution of (\ref{eq:crksk})
that is given by (\ref{eq:trPS}).

\paragraph{Remark:}

An alternative way of writing the invariants $\tau^{(n)}(\ld_a)$ is by
using a tensor $K_a$ given by
\be \label{eq:k}
K_a\,=\,\KK_a S^+_a\,=\,tr_{a\pp}\left\{ P_{a\pp,\hat{a}} S^+_a
\,^{t_a\!}\left((\,^{t_a\!}S^+_{a\pp,\hat{a}})^{-1}\right)
\right\} \   , \ee
leading to the following explicit expression for the invariants
\bea
\tau^{(n)}(\ld_a) &=& tr_a\left( K_a T_a\right)
\,=\,tr_{a,a\pp}\left\{ \PP_{a\pp,\hat{a}}
S^+_a\,^{t_a\!}\left((\,^{t_a\!}S^+_{a\pp,\hat{a}})^{-1}\right)
T^0_{\hat{a}} R^-_a\right\}\     ,\label{eq:trKK} \\
&&\    \ (n=\ell(a)=\ell(a\pp)\ \ ,\ \ \ld_a=\ld_{a\pp} )\  ,  \nn
\eea
and the choice of $\ld_a$ for which $R^-_a$ is a projector. We mention,
finally, that similar
objects have been considered in e.g. \cite{FRT} in the construction of
central elements of quantum groups. However, the connection with exact
invariants of quantum mappings has to our knowledge not been derived
before.

\section{The Gel'fand-Dikii Hierarchy}
\setcounter{equation}{0}

We now present as special examples of the construction given in the
previous sections a specific class of quantum mappings associated with
the lattice Gel'fand-Dikii hierarchy, cf. \cite{NPCQ} which is a
specific class of $N\times N$ matrix lattice models whose continuum
limit reduces to the usual GD hierarchy of equations.
In \cite{NC} the $R,S$-matrix structure for the GD mappings was
introduced, where we established that the full Yang-Baxter structure
(\ref{eq:LLL})-(\ref{eq:MMM}) is verified for these mappings together
with the following solution of the quantum relations
(\ref{eq:RRR},\ref{eq:RSS}) together with (\ref{eq:RS}), namely
\bea
R^+_{12}&=& R^+_{12}(\ld_1,\ld_2)\,=\,R^-_{12}+
\frac{h}{\ld_2}Q_{12} -\frac{h}{\ld_1}Q_{21} \  \ ,  \nn \\
R^-_{12}&=& R^-_{12}(\ld_1,\ld_2)\,=\,{\bf 1} +
h\,\frac{P_{12}}{\ld_1-\ld_2}\   \ ,  \label{eq:Rsol}   \\
S^+_{12}&=& S^+_{12}(\ld_1\ld_2)\,=\,S^-_{21}(\ld_2,\ld_1)\,=\,
{\bf 1} - \frac{h}{\ld_2} Q_{12}\   , \nn
\eea
in which
\be \label{eq:Q}
P_{12}\ =\ \sum_{i,j=1}^N\,E_{i,j}\otimes E_{j,i}\      \ ,
\      \ Q_{12}\ =\ \sum_{i=1}^{N-1}\,E_{N,i}\otimes E_{i,N}\  ,
\ee
the $E_{i,j}$ being the generators of $GL_N$ in the fundamental
representation, i.e.
$(E_{i,j})_{kl}=\dd_{ik}\dd_{jl}$. The special case of $N=2$,
leading to a quantization of the KdV mappings was
presented in \cite{NCP}. The solution (\ref{eq:Rsol}) consists of
the usual $N\times N$ rational $R$-matrix, for which the special
choice $R^-_{12}(\ld_1,\ld_2=\ld_1\pm h)$ leads to a projection
matrix.  In fact, choosing the spectral parameters
according to
\[ \ld_a=(\ld_1,\dots,\ld_n)\    \ ,\   \ \ld_{i+1}\,=\,\ld_i+h\   \ ,
\   \ (i=1,\dots ,n-1)\   , \]
the matrix $R^-_a$ becomes a fully antisymmetric tensor acting
on the $n$-fold tensor product of auxiliary vector spaces $\left(
\Bbb C^N\right)^{\otimes n}$, cf. eq. (\ref{eq:rpp}). \\

To implement the construction of invariants for the mappings in the
GD hierarchy we need the following ingredients.\\
{\bf a)} We consider a periodic chain of $2P$, ($P=1,2,\dots$), sites
labelled by
$n$, and elementary matrices $V_n$ of the form
\be
V_n\ =\ \Lambda_n \left( {\bf 1}\ +\ \sum_{i>j=1}^N\,v_{i,j}(n)E_{i,j}\
\right)   ,  \label{eq:V}
\ee
with
\be
\Lambda_n=\Lambda(\ld_n)\   \ ,\   \
\Lambda(\ld)\ =\ \lambda E_{N,1}\,+\,\sum_{i=1}^{N-1}\,E_{i,i+1}\   \ ,
\   \ \ld_{2n}=\ld\    \ ,\    \ \ld_{2n+1}=\ld + \omega  ,
\label{eq:Lam}
\ee
where
$$ v_{i,j}=0\  \ ,\  \ (i\neq N\,,\,j\neq 1\,,\,i\neq j+1)\   \ ,\   \
v_{i+1,i}(n)=p_{n+1}\   \ ,\  \ (i=2,\dots , N-2)\  , $$
the $p_n$ being constant parameters such that $p_{2n}=p_{2n+2}$,
and where the only
operator-valued entries are $v_{N,j}$ and $v_{i+1,1}$,
($i,j=1,\dots ,N-1$),
where $\omega = (-p_{2n})^N - (-p_{2n+1})^N$,
at each site of the chain. The matrices $V_n$ depends on a different
spectral parameter $\ld$ or $\ld+\omega$ depending on the even or odd
site of the chain.
We impose now for the matrices $V_n$ the commutation relations
\bse  \label{eq:VV}
\bea
V_{n+1,1} S^+_{n,12}\,V_{n,2} &=& V_{n,2} V_{n+1,1}\  ,
\label{eq:VVa} \\
R^+_{n,12}\,V_{n,1} V_{n,2} &=& V_{n,2} V_{n,1}\,R^-_{n,12}\   ,
\label{eq:VVb} \\
V_{n,1} V_{m,2} &=& V_{m,2} V_{n,1}\  \ ,\   \ |n-m|\geq 2,
\label{eq:VVc}
\eea
\ese
in which $S_n^\pm$ and $R_n^\pm$ depend on the local spectral
parameters $\ld_{n,1}$ resp. $\ld_{n,2}$ associated with the
site $n$, i.e. as in (\ref{eq:Rsol}) with $\ld$ replaced by $\ld_n$.
The entries of the matrices $V_n$ do not depend on the spectral
parameter $\ld$, and are hermitean operators $v_{i,j}$
obeying the following Heisenberg type of commutation relations,
($h=i\hbar$),
\be
\left[ v_{i,j}(n)\,,\,v_{k,l}(m) \right]\ =\
h\left( \dd_{n,m+1}\dd_{k,j+1}\dd_{i,N}
\dd_{l,1}\,-\,\dd_{m,n+1}\dd_{i,l+1}\dd_{k,N}\dd_{j,1} \right)\   .
\label{eq:vpl}
\ee
in agreement with eqs. (\ref{eq:VVa}) and (\ref{eq:VVb}).\\
{\bf b)}
With the identification
\be
L_n(\ld)\ =\ V_{2n}(\ld) V_{2n-1}(\ld + \omega )      \ ,\   \label{eq:L}  \ee
we have a solution of the quantum relations (\ref{eq:LLL}), with
and the quantum $M$-matrix is given by
\bea
M_n &=& \Lambda_{2n}\left( {\bf 1}\,+ \,\sum_{i=2}^{N-1}
v^{\prime}_{i,1}(2n-2) E_{i,1}\,+\,
p_{2n}\sum_{j=2}^{N-2} E_{j+1,j}\right. \nn \\
&& \left. +\,\sum_{j=1}^{N-1} v_{N,j}(2n-1) E_{N,j}\,+
\,w(n)E_{N,1}\,+\,(p_{2n}-p_{2n+1})E_{2,1}\right) \   ,   \label{eq:M}
\eea
The corner entry $w(n)$ is determined by the Zakharov-Shabat
equations (\ref{eq:ZS}).
For the matrix $M_n$ we have the commutation relations
\bse  \label{eq:VM}
\bea
M_{n+1,1} S^+_{12} V_{2n,2} &=& V_{2n,2} M_{n+1,1}\   ,
\label{eq:VMa}  \\
V^{\prime}_{2n-1,1} S^+_{12} M_{n,2} &=&
M_{n,2} V^{\prime}_{2n-1,1}   , \label{eq:VMb}  \\
\left[ M_{n,1}\,,\,V_{2n-k,2} \right] &=&
\left[ M_{n+1,1}\,,\,V^{\prime}_{2n-k,2} \right]\,=\,0\    \ ,\    \
(k\neq 2,1,0,-1)\ . \label{eq:VMc}
\eea \ese
{\bf c)} One can work out the ZS equation (\ref{eq:ZS}) to obtain
explicit expressions for the entries of the $M$-matrix in terms of
the entries of the $V_n$.
The result is rather complicated
\bse \label{eq:5} \bea
v^{\prime}_{2,1}(2n-1) &=& v_{2,1}(2n)\  ,   \label{eq:5a}   \\
v^{\prime}_{i,1}(2n-1) &+& p_{2n+1}v^{\prime}_{i-1,1}(2n-1)\ =\
v_{i,1}(2n)\,+\,p_{2n}v_{i-1,1}(2n)   \  ,\  (i=3,\dots ,N-1)\ ,
\label{eq:5b} \\
v^{\prime}_{N,N-1}(2n-1) &=& v_{N,N-1}(2n)\,+\,p_{2n}\,-\,p_{2n+1}\  ,
\label{eq:5c}  \\
v^{\prime}_{N,j}(2n-1) &=& v_{N,j}(2n)   \  ,\  (j=2,\dots ,N-2)\ ,
\label{eq:5d}   \\
v^{\prime}_{N,1}(2n-1) &+& p_{2n+1}v^{\prime}_{N-1,1}(2n-1)\ =\
v_{N,1}(2n)\,+\,p_{2n}v_{N-1,1}(2n)\   ,   \label{eq:5e}   \\
v^{\prime}_{3,1}(2n-2) &-& v_{3,1}(2n-1)\ =\ v_{2,1}(2n)\left[
p_{2n+1}-p_{2n}+v_{2,1}(2n-1)-v^{\prime}_{2,1}(2n-2)\right] \nn \\
&&+\, \left( w(n) - v^{\prime}_{3,1}(2n-2)\right)\dd_{N,3}\  ,
\label{eq:5f}  \\
v^{\prime}_{N,N-2}(2n) &=& v_{N,N-2}(2n+1)\,-\,
p_{2n}v^{\prime}_{N,N-1}(2n)\,+\,p_{2n+1}v_{N,N-1}(2n+1) \nn \\
&&+\, \left( v^{\prime}_{N,N-1}(2n)\,-\,v_{N,N-1}(2n+1)\right)
(p_{2n+1}-v_{N,N-1}(2n)) \nn  \\
&&+\, \left( w(n+1) - v_{3,1}(2n+1)\right)\dd_{N,3}\  ,
\label{eq:5g} \\
v^{\prime}_{i,1}(2n-2) &=& v_{i,1}(2n-1)\,-\,p_{2n+1}\left[
v^{\prime}_{i-1,1}(2n-2)-v_{i-1,1}(2n-1) \right]  \nn \\
&&+\,v_{i-1,1}(2n)\left[ p_{2n+1}-p_{2n}+v_{2,1}(2n-1)-
v^{\prime}_{2,1}(2n-2)\right] \nn \\
&&-\, \left( w(n)-v^{\prime}_{N,1}(2n-2) \right) \dd_{N,i}\     \ ,
\        \  \   \ (i=4,\dots , N)\ ,  \label{eq:5h}  \\
v^{\prime}_{N,j}(2n) &=& v_{N,j}(2n+1)\,-\,p_{2n}v^{\prime}_{N,j+1}(2n)
\,+\,p_{2n+1}v_{N,j+1}(2n+1)\  ,\nn \\
&&-\left( v^{\prime}_{N,N-1}(2n)- v_{N,N-1}(2n+1)\right) v_{N,j+1}(2n)
\,+\, \left( w(n+1)-v_{N,1}(2n+1) \right) \dd_{j,1}\   ,  \nn \\
&&\            \  \     \ (j=1,\dots ,N-3)\   , \label{eq:5j}  \\
\ld_{2n-1}-\ld_{2n}&=& \,\sum_{k=2}^{N-2} v_{N,k}(2n+1)
v_{k+1,1}(2n)\,-\, \sum_{k=2}^{N-2} v^{\prime}_{N,k}(2n)
v^{\prime}_{k+1,1}(2n-1)\, \nn \\
&&+\,\left( w(n+1)-v^{\prime}_{N,1}(2n) \right)
(v_{2,1}(2n)+p_{2n})   \nn \\
&&-\,\left( v^{\prime}_{N,N-1}(2n)- v_{N,N-1}(2n+1)\right)
\left( v_{N,1}(2n) + p_{2n}v_{N,2}(2n)\right)  \nn \\
&&-\,v^{\prime}_{N,N-1}(2n) \left( p_{2n}v_{N-1,1}(2n-2) + p_{2n+1}
v^{\prime}_{N-1,1}(2n-1)\right)   \nn \\
&&+\,p_{2n} \left( p_{2n+1}v_{N,2}(2n+1) - p_{2n}
v^{\prime}_{N,2}(2n)\right)   \label{eq:5k}  \\
\ld_{2n}-\ld_{2n-1}&=& \sum_{k=2}^{N-2} v_{N,k}(2n)v_{k+1,1}(2n-1)
\,-\,\sum_{k=2}^{N-2} v^{\prime}_{N,k}(2n-1)v^{\prime}_{k+1,1}(2n-2)
 \nn \\
&&-\,(v_{N,N-1}(2n)+p_{2n})
\left( w(n)-v_{N,1}(2n-1) \right) \nn \\
&&-\,\left( v_{N,1}(2n)+p_{2n} v_{N,N-1}(2n)\right)
\left( v^{\prime}_{2,1}(2n-2) - v_{2,1}(2n-1) +
p_{2n}-p_{2n+1}\right)  \nn \\
&&-\,p_{2n} p_{2n+1} \left( v^{\prime}_{N-1,1}(2n-2) -
v_{N-1,1}(2n-1)\right)   \label{eq:5l}  \\
v_{N,N-1}(n) &=& p_{n+2}-v_{2,1}(n-2) \   . \label{eq:5m}
\eea \ese
The mapping
as expressed by (\ref{eq:5}) in terms of the operators $v_{i,j}$ is
not very enlightening, and there is a more convenient set of
variables,
$X_n^{(\ar)}$, ($\ar=1,\dots, N-1$, $n=1,\dots,2P$) introduced in
\cite{NC}, cf. also \cite{NPCQ},
in terms of which the mapping seems more natural. However,
the variables $v_{i,j}$ are the natural ones in connection with
the quantum mapping algebra (\ref{eq:LLL})-(\ref{eq:MM}).

{}From the explicit form of (\ref{eq:5}) it can be shown by explicit
calculation that the mapping is symplectic, i.e. it preserves the
basic commutation relations. Therefore, the relations (\ref{eq:VV})
are preserved under the mapping. A more fundamental reason for the
symplecticity is the existence of an action for the entire family
of GD mappings, \cite{NCC}, cf. also \cite{CNP,N} for the KdV case.
A next step is to work out the commutation relations between the
operators $V_n$ and $M_n$, using the explicit expressions that one
obtains for the entries of $M_n$ in terms of $V_n$ and $V^{\prime}_n$.
This is a fairly elaborate, but straightforward calculation, and we
omit the details. (Some details of the calculation are given in
Appendix C). \\
{\bf d)} Having established the full Yang-Baxter structure for the
mapping (\ref{eq:5}), we can use
now the formalism of the previous section, we can
calculate the explicit $K$-matrices that lead to commuting families
of exact quantum invariants.
The monodromy matrix $T(\ld)$ is constructed from the
matrices $V_n$ by
\be
T(\ld)\ \equiv \stackrel{\longleftarrow}{\prod_{n=1}^{2P}}\
V_n(\lambda )\ ,  \label{eq:TV}
\ee
The invariants for the GD hierarchy are given by eq. (\ref{eq:trKK}),
namely
\be \tau^{(n)}(\ld)\,=\,tr_{1,\dots,n}\left( K_{(1,\dots,n)}
T_{(1,\dots,n)}\right)\,=\,
tr_{1,\dots,n}\left( K_{(1,\dots,n)} T^0_{(n,\dots,1)}
P^-_{(1,\dots,n)} \right) \    \  ,\    \
(n=1,\dots ,N), \label{eq:trGD}
\ee
where
\bea
T^0_{(n,\dots ,1)} &=& T_n S^+_{n,n-1} T_{n-1} S^+_{n,n-2}
S^+_{n-1,n-2} T_{n-2} \cdots T_{2} S^+_{n,1} S^+_{n-1,1}\cdots
S^+_{2,1} T_1\    ,    \nn   \\
&&\      \ \ld_i=\ld + (i-1)h\    \ ,\  \ (i=1,\dots,n)\  ,
\label{eq:ttt}
\eea
and where $P^-_{(1,\dots,n)}$ denotes the antisymmetrizer as in
(\ref{eq:anti}).
The $K$-matrix which is obtained from eq. (\ref{eq:trKK})
turns out to factorize due to the nilpotency of
the $S$-matrix of (\ref{eq:Rsol}). In fact, we find
\bea
K_{(1,\dots ,n)} &=& K_1 (S^+_{2,1})^{-1} K_2 (S^+_{3,1})^{-1}
(S^+_{3,2})^{-1} K_3 \cdots K_{n-1} (S^+_{n,1})^{-1} \cdots
(S^+_{n,n-1})^{-1} K_n\    ,    \label{eq:kkk}   \\
K_i(\ld_i)&=& {\bf 1}\,+\,\frac{h}{\ld_i} Q_{i,i}\,=\,{\bf 1}+(N-1)
\frac{h}{\ld_i} (E_{N,N})_i\    \ ,\    \
\ld_i=\ld + (i-1)h\    ,  \nn
\eea
in which $Q_{i,i}$ denotes the contraction of the tensor $Q$ of (\ref{eq:Q}).
The actual invariants are now obtained by expanding in powers of
the spectral parameter $\ld$. Due to the form of the matrices $V_n$,
(\ref{eq:V}), the monodromy matrix (\ref{eq:TV}) has coefficients
which are simply polynomial in $\ld$. It is a matter of counting
to verify that by considering all invariants thus obtained from
(\ref{eq:trGD}) together with (\ref{eq:kkk}), for $n=1,\dots,N$,
will yield a sufficiently
large family of commuting invariants of the quantum mapping in terms
of the reduced variables $X_n^{(\ar)}$ introduced in \cite{NPCQ,NC}.
We shall not occupy ourselves here with this problem, as we
enviseage to deal with that issue at a later date when we plan to
investigate the representation theory for the mapping algebra,
cf. \cite{NCC}.
The Casimirs for the mapping algebra ${\cal A}$ are given by
the coefficients of the deformed quantum determinant, cf. also e.g.
\cite{Ols}, which is obtained from (\ref{eq:trGD}) for the top value,
$n=N$.

\paragraph{Remark:}

The role played by the `pivot' matrix $\Lambda(u)$ in the
above construction of the GD-hierarchy can be made clear by the
following. Due to the identities
$$ R^+_{12}\ =\ \Lambda_1\,\Lambda_2\,R_{12}\,\Lambda^{-1}_1\,
\Lambda^{-1}_2\    \ ,\    \ S^+_{12}\ =\ \Lambda_2\,S_{12}\,
\Lambda^{-1}_2\    ,  $$
in which
$$R_{12}\equiv R_{12}^-\    \ ,\   \
S_{12}\equiv \Lambda_2^{-1}S^+_{12}\ \Lambda_2  , $$
it is evident that in this case it is not strictly necessary to
introduce two different $R$-matrices $R^{\pm}$. In fact, taking
$R_{12}\equiv R^-_{12}$
we can go over to a `symmetric' version of the quantum relations
(\ref{eq:RLL}), namely
\be
R_{12}\,{\cal L}_{n,1} {\cal L}_{n,2}\ =\
{\cal L}_{n,2} {\cal L}_{n,1}\,R_{12}\    \ ,
\     \  {\cal L}_{n+1,1} S_{12}{\cal L}_{n,2}\ =\
{\cal L}_{n,2} {\cal L}_{n+1,1} \   ,  \label{eq:LLR}
\ee
for ${\cal L}\equiv \Lambda^{-1}\,L$.
However, working with this symmetrized version of the basic equations
has the disadvantage that the relations
(\ref{eq:RS}) and the definition of the monodromy matrix become less
natural. The relation (\ref{eq:RS}) reduces then to
$$ R_{12}\,\Lambda_1\,S_{21}\,\Lambda_2\ =\ \Lambda_2\,S_{12}\Lambda_1
\,R_{12}\     . $$
In the case of the GD-hierarchy the symmetric $R$- and $S$-matrices
are given by
\be
R_{12}\ =\ {\bf 1}\ +\ h\,\frac{P_{12}}{\lambda_1\,-\,\lambda_2}\
         \ ,\       \
S_{12}\ =\ {\bf 1}\ -\ h\,\sum_{i=1}^{N-1}\,E_{N,i}\otimes E_{i+1,1}\   .
\ee

\section{Conclusions}

We have given a general construction of exact quantum invariants
associated with the quantum mapping algebra given by eqs.
(\ref{eq:LLL})-(\ref{eq:MM}). The relevant algebra for the monodromy
matrix is given by the equations (\ref{eq:RTST}) together with
(\ref{eq:MMR}) and (\ref{eq:TMSM}). The construction of commuting
families of quantum operators follows basically the same philosophy
as the construction in \cite{Skly}. As a consequence, weak
integrability of the mappings (according to the terminology introduced
in \cite{NC}) is established by
developing the fusion algebra associated with the mapping algebra,
which ensures that commuting families of (not necessarily invariant)
operators are mapped to again commuting families.
However, we have shown under what conditions
the statement of strong integrability of mappings can also
be made for the algebra ${\cal A}$ of the quantum mappings.
It turns out
that there is actually a unique family of
commuting operators made up of exact quantum {\em invariants} of
the mapping. Implementing the fusion procedure we have
established that this invariance can be pushed
to the level of higher-order quantum minors and determinants,
associated with e.g. higher-spin models or (as we have demonstrated
by the example of the GD hierarchy) $N\times N$ matrix quantum models.
These results now open the way to the exact `diagonalization' of the
quantum discrete-time evolution, e.g. via the algebraic Bethe's
Ansatz, \cite{KIB}, applied to the quantum mapping algebra.

We have not embarked in this paper on issues of representation theory,
in terms of which the results presented here can no doubt be generalized.
Another  issue that we have not addressed here is
the construction of the generating
quantum operator of the discrete-time flow. A construction of such
operators has been given recently in \cite{FadSal} in the context of
the Quantum Volterra model.
However, for the mappings of the GD hierarchy, a direct connection
between the
unitary operator of the quantum canonical transformation, as was
derived in \cite{N}, and the monodromy matrix of the system, has still
to be established.
A direct construction of these operators on the basis of quantum
actions for the mappings is under investigation, \cite{NCC}.

We finish by mentioning in this context also the recent interest in
$q$-difference analogues of the Knizhnik-Zamolodchikov equations, that
have appeared recently in connection with the representation theory
of affine quantum groups, cf. e.g. \cite{Smirn,FR}. This is yet
another manifestation of the inherent discrete nature of quantum
groups and related objects. It would be of interest to bring all these
aspects together in one global difference approach to the underlying
structures.

\subsection*{Acknowledgement}

FWN is grateful to Profs. L.D. Faddeev and L.A. Takhtajan for
stimulating discussions.
\pagebreak

\subsection*{Appendix A}
\setcounter{equation}{0}
\def\theequation{A.\arabic{equation}}

In this appendix we derive the relations (\ref{eq:Rr}), (\ref{eq:SR})
and (\ref{eq:sr}).
The proof of eq. (\ref{eq:rr}) are obtained by direct iteration
of the Yang-Baxter equations (\ref{eq:RR}). \\

\noindent
{\bf i)}
Eqs. (\ref{eq:Rr}) can be proven by induction. They hold trivially for
the case $\ell(a)=1$. For the induction procedure it suffices to show
how the equations  can be build up if we merge multi-indices, thus
showing that from smaller multi-indices we can construct the same
statements for multi-indices made up of the smaller pieces. (This will
be the philosophy for all the proofs). Thus, assuming that
(\ref{eq:rrc}) hold for multi-indices $a,b,\dots $, we show that they also
hold for merged indices such as $(ab)$, etc. For instance,
if we want to demonstrate (\ref{eq:rrc}), we perform the following
sequence of steps
\bea
R^{\pm}_{(ab)} S^{\pm}_{c,(ab)} &=& R^{\pm}_{\hat{a},b} R^{\pm}_a
R^{\pm}_b S^{\pm}_{c,a} S^{\pm}_{c,b}     \nn \\
&=& R^{\pm}_{\hat{a},b} S^{\pm}_{c,\hat{a}} S^{\pm}_{c,\hat{b}}
R^{\pm}_a R^{\pm}_b \nn \\
&=& S^{\pm}_{c,\hat{b}} S^{\pm}_{c,\hat{a}} R^{\pm}_{\hat{a},b}
R^{\pm}_a R^{\pm}_b \nn \\
&=& S^{\pm}_{c,\widehat{(ab)}} R^{\pm}_{(ab)}\     \ ,\   \
\widehat{(ab)}=(\hat{b}\hat{a})\   ,   \label{eq:A1}
\eea
and similarly for the other relations. \\

\noindent
{\bf ii)} Eq. (\ref{eq:SR}) can again be simply proven by induction. Eq.
(\ref{eq:SR}) clearly holds for $\ell(a)=1$, in which case the objects
$R^{\pm}_a$ and $S^{\pm}_a$ are simply the unit matrices acting on
one single vector space. Now, assuming that
(\ref{eq:SR}) holds for a fixed value of $\ell(a)$, then we can iterate
as follows
\bea
R^{-}_{\hat{a},(bc)} S^{+}_{(bca)} &=&
R^{-}_{\hat{a},b} R^{-}_{\hat{a},c} S^{+}_{(ca)}
S^{+}_{\hat{b},\widehat{(ca)}} S^{+}_b  \nn \\
&=& R^{-}_{\hat{a},b} S^{+}_{(ac)} R^{+}_{\hat{a},c}
S^{+}_{\hat{b},\hat{a}} S^{+}_{\hat{b},\hat{c}} S^{+}_b   \nn \\
&=& R^{-}_{\hat{a},b} S^{+}_c S^{+}_{\hat{a},\hat{c}} S^{+}_a
S^{+}_{\hat{b},\hat{c}} S^{+}_{\hat{b},\hat{a}} S^{+}_b R^{+}_{\hat{a},c}   \nn
\\
&=& S^{+}_c S^{-}_{\hat{c},\hat{b}} S^{-}_{\hat{c},\hat{a}}
R^{-}_{\hat{a},b} S^{+}_a S^{+}_{\hat{b},\hat{a}} S^{+}_b
R^{+}_{\hat{a},c}  \nn \\
&=& S^{+}_c S^{+}_{\hat{b},\hat{c}} S^{+}_{\hat{a},\hat{c}} S^{+}_{(ab)}
R^{+}_{\hat{a},b} R^{+}_{\hat{a},c}\,=\, S^{+}_{(abc)}
R^{+}_{\hat{a},(bc)}\   .  \label{eq:A2}
\eea
where in the first and last step we have used the decomposition of
$S^{\pm}_{(abc)}$ according to (\ref{eq:A}),
as well as the induction assumption, and in the second and third step
the fused Yang-Baxter relations (\ref{eq:rss}). A similar line of
steps leads to
\be
R^{-}_{\widehat{(ab)},c} S^{+}_{(cab)}\,=\,
S^{+}_{(abc)} R^{+}_{\widehat{(ab)},c}\   ,  \label{eq:A3}
\ee
thus  we can break down the multi-indices in parts repeating the
relations (\ref{eq:A1}) and (\ref{eq:A2}) in successive steps. \\

\noindent
{\bf iii)} In order to prove eqs. (\ref{eq:sr}), we break down the
multi-index relation as follows
\bea
S^{+}_{(ab)} R^{+}_{(ab)} &=&
S^{+}_b S^{+}_{\hat{a},\hat{b}} S^{+}_a
R^{+}_b R^{+}_{\hat{a},\hat{b}} R^{+}_a \nn  \\
&=& S^{+}_b R^{+}_b  S^{+}_{\hat{a},b} S^{+}_a
R^{+}_{\hat{a},\hat{b}} R^{+}_a \nn  \\
&=& R^{-}_b S^{+}_{\hat{b}} S^{+}_{\hat{a},b} S^{+}_a
R^{+}_{\hat{a},\hat{b}} R^{+}_a \nn  \\
&=& R^{-}_b S^{+}_{(a\hat{b})} R^{+}_{\hat{a},\hat{b}} R^{+}_a \nn \\
&=& R^{-}_b R^{-}_{\hat{a},\hat{b}} S^{+}_{(\hat{b}a)} R^{+}_a \nn \\
&=& R^{-}_b R^{-}_{\hat{a},\hat{b}} S^{+}_a S^{+}_{b,\hat{a}}
S^{+}_{\hat{b}} R^{+}_a \nn  \\
&=& R^{-}_b R^{-}_{\hat{a},\hat{b}} R^{-}_a S^{+}_{\hat{a}}
S^{+}_{b,a} S^{+}_{\hat{b}} \nn  \\
&=& R^{-}_{(ab)} S^{+}_{(\hat{b}\hat{a})}\,=\,
R^{-}_{(ab)} S^{+}_{\widehat{(ab)}} \  ,   \label{eq:A4}
\eea
where use has been made of the decomposition of $S^{\pm}_{(ab)}$ in
the first, fourth and last step,
the relation (\ref{eq:rrc}) in
the second step and fore-last step and the induction assumption in the
other step.

\renewcommand{\pp}{^{\prime}}

\subsection*{Appendix B}
\setcounter{equation}{0}
\def\theequation{B.\arabic{equation}}

In this appendix, we give details of the verification that
(\ref{eq:trPS}) obeys eq. (\ref{eq:crksk}) when multiplied by the
projectors $R^-_a$ and $R^-_b$ at both sides.
In fact, inserting $\KK_a$ into the
right-hand side of (\ref{eq:crksk}) we perform the following
sequence of manipulations
\bea
\KK_b\,^{t_b\!}&& \left( (\,^{t_b\!}\SS^+_{a,\hat{b}})^{-1}\right)
\KK_a\, ( \RR^+_{\hat{a},b})^{-1} \,=    \nn \\
&&=\,tr_{a\pp,b\pp}\left\{ \PP_{b\pp,\hat{b}}
\,^{t_{b\pp}\!}\left( (\,^{t_{b\pp}\!}\SS^+_{b\pp,\hat{b}})^{-1}
\right) \,^{t_a\!}\left( (\,^{t_a\!}\SS^+_{a,\hat{b}})^{-1} \right)
\PP_{a\pp,\hat{a}}
\,^{t_{a\pp}\!}\left( (\,^{t_{a\pp}\!}\SS^+_{a\pp,\hat{a}})^{-1}
\right) ( \RR^+_{\hat{a},b})^{-1} \right\}   \nn \\
&&=\,tr_{a\pp,b\pp}\left\{ \PP_{a\pp,\hat{a}}\PP_{b\pp,\hat{b}}
\ ^{t_{a\pp b\pp}\!} \left[ (\,^{t_{b\pp}\!}\SS^+_{b\pp,\hat{b}})^{-1}
(\,^{t_{a\pp}\!}\SS^+_{a\pp,\hat{a}})^{-1}
(\,^{t_{a\pp}\!}\SS^+_{a\pp,\hat{b}})^{-1}
( \RR^+_{\hat{a},b})^{-1} \right] \right\}   \nn \\
&&=\,tr_{a\pp,b\pp}\left\{ \PP_{a\pp,\hat{a}}\PP_{b\pp,\hat{b}}
\ ^{t_{a\pp b\pp}\!} \left[ (\,^{t_{b\pp}\!}\SS^+_{b\pp,\hat{b}})^{-1}
( \RR^+_{\hat{a},b})^{-1}
(\,^{t_{a\pp}\!}\SS^+_{a\pp,\hat{b}})^{-1}
(\,^{t_{a\pp}\!}\SS^+_{a\pp,\hat{a}})^{-1} \right] \right\}   \nn \\
&&=\,tr_{a\pp,b\pp}\left\{ \,^{t_{a\pp}\!}\PP_{a\pp,\hat{a}}
\,^{t_{b\pp}\!}\PP_{b\pp,\hat{b}} \,^{t_{b\pp}\!}\SS^+_{b\pp,\hat{a}}
\left[ \left( \,^{t_{b\pp}\!}\SS^+_{b\pp,\hat{a}}\right)^{-1}
(\,^{t_{b\pp}\!}\SS^+_{b\pp,\hat{b}})^{-1}
( \RR^+_{\hat{a},b})^{-1} \right]
(\,^{t_{a\pp}\!}\SS^+_{a\pp,\hat{b}})^{-1}
(\,^{t_{a\pp}\!}\SS^+_{a\pp,\hat{a}})^{-1}  \right\}   \nn \\
&&=\,tr_{a\pp,b\pp}\left\{ \,^{t_{a\pp}\!}\PP_{a\pp,\hat{a}}
\,^{t_{b\pp}\!}\PP_{b\pp,\hat{b}} \,^{t_{b\pp}\!}\SS^+_{b\pp,\hat{a}}
( \RR^+_{\hat{a},b})^{-1}
(\,^{t_{b\pp}\!}\SS^+_{b\pp,\hat{b}})^{-1}
(\,^{t_{b\pp}\!}\SS^+_{b\pp,\hat{a}})^{-1}
(\,^{t_{a\pp}\!}\SS^+_{a\pp,\hat{b}})^{-1}
(\,^{t_{a\pp}\!}\SS^+_{a\pp,\hat{a}})^{-1} \right\}\  .  \label{eq:B1}
\eea
Multiplying at this point (\ref{eq:B1}) at the right by
$R^-_aR^-_b$, we can use (\ref{eq:csr}) to move these through the
matrices $\SS$. Next we multiply (\ref{eq:B1}) at the left by
the same factors and use the relations
\be
\,^{t_{a\pp}\!}\PP_{a,\hat{a}} X_a\,=\,^{t_{a\pp}\!}\PP_{a,\hat{a}}
\,^{t_{a\pp}\!}X_{a\pp}\    \ , \
\ X_a\,^{t_{a\pp}\!}\PP_{a,\hat{a}}\,=
\,^{t_{a\pp}\!}X_{a\pp}\,^{t_a\!}\PP_{a,\hat{b}} \   , \label{eq:B2}
\ee
which are a consequence of (\ref{eq:pp}), to transpose $R^-_a$ and
$R^-_b$ into $\,^{t_{a\pp}\!}R^-_{a\pp}$ and
$\,^{t_{b\pp}\!}R^-_{b\pp}$, and bring them under the trace. Then
applying the cyclicity of the double trace we move these factors
to the right and, by using again
(\ref{eq:csr}), through the matrices $\SS$.
In this way we find
\bea
R^-_a && R^-_b \KK_b\,^{t_b\!} \left( (\,^{t_b\!}\SS^+_{a,\hat{b}})^{-1}\right)
\KK_a\, ( \RR^+_{\hat{a},b})^{-1} R^-_a R^-_b \,=    \nn \\
&&=\,tr_{a\pp,b\pp}\left\{ \,^{t_{a\pp}\!}\PP_{a\pp,\hat{a}}
\,^{t_{b\pp}\!}\PP_{b\pp,\hat{b}} \,^{t_{b\pp}\!}\SS^+_{b\pp,\hat{a}}
( \RR^+_{\hat{a},b})^{-1} R^-_a R^-_b
\,^{t_{a\pp}\!}R^-_{a\pp} \,^{t_{b\pp}\!}R^-_{b\pp}
(\,^{t_{b\pp}\!}\SS^+_{\hat{b\pp},b})^{-1}
(\,^{t_{b\pp}\!}\SS^+_{\hat{b\pp},a})^{-1}
(\,^{t_{a\pp}\!}\SS^+_{\hat{a\pp},b})^{-1}
(\,^{t_{a\pp}\!}\SS^+_{\hat{a\pp},a})^{-1} \right\}\  .  \nn \\
\label{eq:B3}
\eea
Then, we can use
again eqs. (\ref{eq:B2}), and subsequently (\ref{eq:csr}), the
Yang-Baxter equations and (\ref{eq:crs}) together with
the property that $R^-_a$ and $R^-_b$ are projectors,
to transmute the combination
\bea
\,^{t_{a\pp}\!}\PP_{a\pp,\hat{a}}&&\,^{t_{b\pp}\!}\PP_{b\pp,\hat{b}}
\,^{t_{b\pp}\!}\SS^+_{b\pp,\hat{a}} ( \RR^+_{\hat{a},b})^{-1} R^-_a
R^-_b \,^{t_{a\pp}\!}R^-_{a\pp}\,^{t_{b\pp}\!}R^-_{b\pp}   \nn \\
&&=\,
\,^{t_{a\pp}\!}\PP_{a\pp,\hat{a}} \,^{t_{b\pp}\!}\PP_{b\pp,\hat{b}}
R^-_a R^-_b\,\SS^+_{b,\hat{a}} ( \RR^+_{\hat{a},b} )^{-1} R^-_a
R^-_b   \nn \\
&&=\,^{t_{a\pp}\!}\PP_{a\pp,\hat{a}} \,^{t_{b\pp}\!}\PP_{b\pp,\hat{b}}
R^-_a R^-_b\,\SS^+_{\hat{b},\hat{a}} R^-_a R^-_b
( \RR^+_{a,\hat{b}} )^{-1} R^-_a
R^-_b   \nn \\
&&=\,^{t_{a\pp}\!}\PP_{a\pp,\hat{a}} \,^{t_{b\pp}\!}\PP_{b\pp,\hat{b}}
R^-_a R^-_b\,\SS^+_{\hat{b},\hat{a}} ( \RR^+_{\hat{a},b} )^{-1} R^-_a
R^-_b   \nn \\
&&=\,^{t_{a\pp}\!}\PP_{a\pp,\hat{a}} \,^{t_{b\pp}\!}\PP_{b\pp,\hat{b}}
R^-_a R^-_b\,( \RR^-_{\hat{a},b})^{-1} \SS^+_{\hat{a},\hat{b}} R^-_a
R^-_b   \nn \\
&&=\,^{t_{a\pp}\!}\PP_{a\pp,\hat{a}} \,^{t_{b\pp}\!}\PP_{b\pp,\hat{b}}
R^-_a R^-_b\, ( \RR^-_{a,\hat{b}} )^{-1} R^-_a R^-_b
\SS^+_{a,\hat{b}} R^-_a R^-_b
 \nn \\
&&=\,^{t_{a\pp}\!}\PP_{a\pp,\hat{a}} \,^{t_{b\pp}\!}\PP_{b\pp,\hat{b}}
\,^{t_{a\pp}\!}\SS^+_{a\pp,\hat{b}}
\,^{t_{a\pp b\pp}\!}( \RR^-_{a\pp,\hat{b\pp}})^{-1}
R^-_a R^-_b \,^{t_{a\pp}\!}R^-_{a\pp}\,^{t_{a\pp}\!}R^-_{a\pp}
\    .  \label{eq:B4}
\eea
Using then again (\ref{eq:csr}) and cyclicity of the trace to move the
factors $R^-_a$, $R^-_b$, $\,^{t_{a\pp}\!}R^-_{a\pp}$ and
$\,^{t_{a\pp}\!}R^-_{a\pp}$ to the right and out of the trace again,
we obtain from  (\ref{eq:B3})
\bea
&&R^-_a R^-_b tr_{a\pp,b\pp}\left\{ \,^{t_{a\pp}\!}\PP_{a\pp,\hat{a}}
\,^{t_{b\pp}\!}\PP_{b\pp,\hat{b}} \,^{t_{a\pp}\!}\SS^+_{a\pp,\hat{b}}
(\,^{t_{a\pp b\pp}\!}\RR^-_{a\pp,\hat{b\pp}})^{-1}
(\,^{t_{b\pp}\!}\SS^+_{b\pp,\hat{b}})^{-1}
(\,^{t_{a\pp}\!}\SS^+_{a\pp,\hat{b}})^{-1}
(\,^{t_{b\pp}\!}\SS^+_{b\pp,\hat{a}})^{-1}
(\,^{t_{a\pp}\!}\SS^+_{a\pp,\hat{a}})^{-1} \right\}  R^-_a R^-_b \nn \\
&&=\,R^-_a R^-_b tr_{a\pp,b\pp}\left\{ \,^{t_{a\pp}\!}\PP_{a\pp,\hat{a}}
\,^{t_{b\pp}\!}\PP_{b\pp,\hat{b}}
(\,^{t_{b\pp}\!}\SS^+_{b\pp,\hat{b}})^{-1}
(\,^{t_{a\pp}\!}\SS^+_{a\pp,\hat{a}})^{-1}
(\,^{t_{b\pp}\!}\SS^+_{b\pp,\hat{a}})^{-1}
(\,^{t_{a\pp b\pp}\!}\RR^-_{a\pp,\hat{b\pp}})^{-1}
 \right\} R^-_a R^-_b  \nn  \\
&&=\,R^-_a R^-_b tr_{a\pp,b\pp}\left\{ \PP_{a\pp,\hat{a}}
\PP_{b\pp,\hat{b}} (\RR^-_{a\pp,\hat{b\pp}})^{-1}
\,^{t_{a\pp}\!}\left( (\,^{t_{a\pp}\!}\SS^+_{a\pp,\hat{a}})^{-1}\right)
\,^{t_{b\pp}\!}\left( (\,^{t_{b\pp}\!}\SS^+_{b\pp,\hat{a}})^{-1}\right)
\,^{t_{b\pp}\!}\left( (\,^{t_{b\pp}\!}\SS^+_{b\pp,\hat{b}})^{-1}\right)
 \right\} R^-_a R^-_b  \nn  \\
&&=\, R^-_a R^-_b (\RR^-_{a,\hat{b}})^{-1}
tr_{a\pp}\left\{ \PP_{a\pp,\hat{a}}
\,^{t_a\!}\left( (\,^{t_a\!}\SS^+_{a\pp,\hat{a}})^{-1} \right)
 \right\} \,
\,^{t_b\!}\left( (\,^{t_b\!}\SS^+_{b,\hat{a}})^{-1} \right)
tr_{b\pp}\left\{ \PP_{b\pp,\hat{b}}
\,^{t_b\!}\left( (\,^{t_b\!}\SS^+_{b\pp,\hat{b}})^{-1} \right)
 \right\} R^-_a R^-_b  \nn   \\
&&=\,R^-_a R^-_b\left( \RR^-_{a,\hat{b}}\right)^{-1} \KK_a
\,^{t_a\!}\left( (\,^{t_a\!}\SS^-_{\hat{a},b})^{-1}\right)
\KK_b\ R^-_a R^-_b  ,
\eea
leading to the conclusion of Proposition 3.

\subsection*{Appendix C}
\setcounter{equation}{0}
\def\theequation{C.\arabic{equation}}

In this appendix we discuss the commutation relations for the matrices
$M_n$ as given by eqs. (\ref{eq:VM}) and (\ref{eq:MMM}).
{}From the explicit form of the mapping (\ref{eq:5}) one observes the
following:
\bse \label{eq:C1}  \bea
w(n)-v_{N,1}(2n-1)&=& \Rrr A\left( v\pp_{2,1}(2n-2)-v_{2,1}(2n-1)\,,
\,\{ v_{j,1}(2n)\}_{j=2,\dots,N-1} \right) \  ,  \label{eq:C1a} \\
w(n+1)-v\pp_{N,1}(2n)&=& \Rrr B\left( v\pp_{N,N-1}(2n)-v_{N,N-1}(2n+1)
\,,\, \{ v_{N,j}(2n),v_{N,j}(2n+1)\}_{j=2,\dots,N-1}
\right)    \   , \nn \\
\label{eq:C1b}
\eea \ese
in which $\Rrr A$ and $\Rrr B$ are some functions that can be
explicitely inferred by solving $v\pp_{i,1}(2n-2)-
v_{i,1}(2n-1)$ and $w(n)-v_{N,1}(2n-1)$ iteratively from
(\ref{eq:5f}) and (\ref{eq:5h}) and solving $v\pp_{N,j}(2n)-
v_{N,j+1}(2n+1)$ and $v\pp_{N,1}(2n)-w(n+1)$ from
(\ref{eq:5j}). From  (\ref{eq:5k}) and (\ref{eq:5l}),
using the relations obtained for
$v\pp_{i,1}(2n-2)-v_{i,1}(2n-1)$ and $v\pp_{N,j}(2n)-v_{N,j+1}(2n+1)$,
we find
\bse \label{eq:C2}  \bea
w(n)-v_{N,1}(2n-1)&=& \Rrr C\left( v\pp_{2,1}(2n-2)-v_{2,1}(2n-1)\,,\,
V_{2n}, \{ v_{j,1}(2n-1)\}_{j=3,\dots,N-1} \right)\  ,   \label{eq:C2a} \\
w(n+1)-v\pp_{N,1}(2n)&=& \Rrr D\left( v\pp_{N,N-1}(2n)-v_{N,N-1}(2n+1)
\,,\,
V_{2n}\,,\,\{ v_{N,j}(2n),v_{N,j}(2n+1)\}_{j=2,\dots,N-1}
\right)    \   , \nn \\
\label{eq:C2b}
\eea \ese
with explicit forms for $\Rrr C$ and $\Rrr D$, which we do not
specify. From
(\ref{eq:C1b}) and (\ref{eq:C2b}) one can solve
\be \label{eq:C3}
v\pp_{N,N-1}(2n)-v_{N,N-1}(2n+1)\,=\, \Rrr E\left( V_{2n}\,,\,
\{ v_{N,j}(2n+1)\}_{j=2,\dots,N-1}\right) \  ,
\ee
and from (\ref{eq:C1a}) and (\ref{eq:C2a}) it follows that
\be \label{eq:C4}
v\pp_{2,1}(2n-2)-v_{2,1}(2n-1)\,=\, \Rrr F \left( V_{2n}\,,\,
\{ v_{j,1}(2n-1)\}_{j=3,\dots,N-1}\right) \  ,
\ee
$\Rrr E$ and $\Rrr F$ denoting some explicit expressions of the given
arguments.
Using (\ref{eq:5m}) in combination with (\ref{eq:C3}) and
(\ref{eq:C4}) it follows that
\be \label{eq:C5}
v\pp_{2,1}(2n-2)-v_{2,1}(2n-1)\,=\,v\pp_{N,N-1}(2n)-v_{N,N-1}(2n+1)
\,=\,\Rrr G \left( V_{2n} \right)\  ,
\ee
depending only on a given function $\Rrr G$ of $V_{2n}$.
Inserting (\ref{eq:C5}) into (\ref{eq:5f}) and (\ref{eq:5h}) we
immediately obtain the relations
\be M_n-V_{2n-1}\,=\,\Rrr H \left( V_{2n}\right)\   ,  \label{eq:C6}
\ee
also depending only on $V_{2n}$,
implying that $M_n$ commutes with $V_{2n-j}$ for $j\neq 2,1,0,-1$ as in
eq. (\ref{eq:VMc}). Eq. (\ref{eq:VMa}) follows from the
commutation relation $[ M_{n+1}-V_{2n+1} \stackrel{\otimes}{,} V_{2n} ]=0$,
as in eq. (\ref{eq:VVa}). Furthermore, from (\ref{eq:C5}),
(\ref{eq:5g}) and (\ref{eq:5j}) we have a relation of the form
\be V\pp_{2n}-M_{n+1}\,=\,\Rrr I \left( V_{2n},
\{ v_{N,j}(2n+1)\}_{j=2,\dots,N-1}\right) \  . \label{eq:C7}
\ee
{}From the fact that all elements of $V\pp_{2n+1}$ can be expressed in
terms of $V_{2n+2}$, cf. eqs. (\ref{eq:5a})-(\ref{eq:5e}), we
have that $[ M_{n+1}-V\pp_{2n} \stackrel{\otimes}{,} V\pp_{2n+1} ]=0$,
and eq. (\ref{eq:VMb}) follows taking into account that the mapping
is symplectic.

Considering that from (\ref{eq:C6}) and (\ref{eq:C7}) we have
\be V\pp_{2n}-V_{2n+1}\,=\,\Rrr H \left( V_{2n+2}\right)
\,+\,\Rrr I \left( V_{2n},\{ v_{N,j}(2n+1)\}_{j=2,\dots,N-1}\right) \  .
\label{eq:C8}
\ee
it is seen that from all the $V\pp_{2n-2k}$, only $V\pp_{2n}$ and
$V\pp_{2n+1}$ can give rise to terms involving $V_{2n-1}$, $V_{2n}$ and
$V_{2n+1}$ that have nonvanishing commutation relations with
$V\pp_{2n}-M_{n+1}$. Hence
$[ M_{n+1} \stackrel{\otimes}{,} V\pp_{2n-k} ]=0$, for $k\neq 2,1,0,-1$,
as in (\ref{eq:VMc}). From (\ref{eq:C6}) we have
\be
[ M_{n} , v_{N,j}(2n+1) ]\,=\,0\   \ ,\   \
(j=2,\dots , N-1)\ , \label{eq:C9}
\ee
Considering the relation
\be v\pp_{i,1}(2n-2)-v_{i,1}(2n-1)\,=\,\Rrr J \left( V_{2n}\right)\   ,
\ee
which follows from (\ref{eq:C5}) in combination with (\ref{eq:5f})
and (\ref{eq:5h}), it follows that $v\pp_{i,1}(2n-1)$ can only have
nonvanishing commutation relations with $V_{2n+1}$, $V_{2n}$ and
$v_{i,1}(2n-1)$. Taking into account (\ref{eq:C7}) and the symplecticity
of the mapping, we find
\be
[ M_{n} , v\pp_{i,1}(2n-2) ]\,=\,0\   \ ,\   \
(j=2,\dots , N-1)\ . \label{eq:C10}
\ee
Eq. (\ref{eq:C9}) and (\ref{eq:C10}) can be combined to yield that
$ [ M_{n} \stackrel{\otimes}{,} M_n ]\,=\,0$ , from which eq.
(\ref{eq:RMM})
can be derived in analogy with eq. (\ref{eq:VVb}) for the matrix $V_n$.
Finally, from the updated (`primed') version of eq. (\ref{eq:C6}) and
the commutation relation $ [ V\pp_{2n} \stackrel{\otimes}{,} M_n ]\,=\,0$ ,
we obtain
\be \label{eq:C11}
[ M\pp_{n}-V\pp_{2n-1} \stackrel{\otimes}{,} M_n ]=0\   ,
\ee
leading with eq. (\ref{eq:VMb}) to eq. (\ref{eq:MSM}).
This proves all the
commutation relations between the matrices $V_n$ and $M_n$ as given
in section 5.
\pagebreak


\begin{thebibliography}{99}
\bibitem{F}
L.D. Faddeev, in {\em D\'eveloppements R\'ecents en Th\'eorie des
Champs et M\'ecanique Statistique}, eds. J.-B. Zuber and R. Stora,
(North-Holland Publ. Co., 1984), p.561.
\bibitem{KIB}
V.E. Korepin, A.G. Izergin and N.M. Bogoliubov, {\em Quantum Inverse
Scattering Method and Correlation Functions}, (Cambridge University
Press, 1992).
\bibitem{KR0}
P.P. Kulish and N.Yu. Reshetikhin, J. Sov. Math. {\bf 23} (1983) 2435.
\bibitem{Jim}
M. Jimbo, Lett. Math.
Phys. {\bf 10} (1985) 63, ibid. {\bf 11} (1986) 247, Commun. Math. Phys.
{\bf 102} (1986) 537.
\bibitem{Drin}
V.G. Drinfel'd, {\em Quantum Groups}, Proc. ICM Berkeley 1986, ed.
A.M. Gleason, (AMS, Providence, 1987), p. 798.
\bibitem{Drin2}
V.G. Drinfel'd, Sov. Math. Dokl. {\bf 32} (1985) 245; ibid. {\bf 36}
(1988) 212.
\bibitem{FRT}
L.D. Faddeev, N.Yu. Reshetikhin and L.A. Takhtadzhyan, Leningrad
Mathematical J. [now: St. Petersburg Math. J.] {\bf 1} (1990) 193.
\bibitem{FF}
L.D. Faddeev, Lectures at the Carg\`ese Summer School on "New
Symmetries in Quantum Field Theory", Preprint HU-TFT-92-5.
\bibitem{Vol}
A.Yu. Volkov, {\em Quantum Volterra Model}, Preprint HU-TFT-92-6.
\bibitem{Smir}
F.A. Smirnov, {\em Form Factors in Completely Integrable Models of
Quantum Field Theory}, in Adv. Ser. Math. Phys. {\bf 14} (World
Scientific, Signapore, 1992).
\bibitem{NPC}
F.W. Nijhoff, V.G. Papageorgiou and H.W. Capel, in Proc. of the
Intl. Workshop on Quantum Groups, The Euler Intl. Math. institute,
Leningrad, ed. P.P. Kulish, (Springer Lecture Notes Math. {\bf 1510},
1991), p. 312.
\bibitem{NPCQ}
F.W. Nijhoff, V.G. Papageorgiou, H.W. Capel and G.R.W. Quispel, Inv.
Probl. {\bf 8} (1992) 597.
\bibitem{PNC}
V.G. Papageorgiou, F.W. Nijhoff and H.W. Capel, Phys. Lett. {\bf 147A}
(1990) 106.
\bibitem{CNP}
H.W. Capel, F.W. Nijhoff and V.G. Papageorgiou, Phys.
Lett. {\bf 155A} (1991) 377.
\bibitem{Ves}
A.P. Veselov, Russ. Math. Surv. {\bf 46} (1991) 1-51.
\bibitem{NCP}
F.W. Nijhoff, H.W. Capel and V.G. Papageorgiou, Phys. Rev. {\bf A46}
(1992) 2155.
\bibitem{NC}
F.W. Nijhoff and H.W. Capel, Phys. Lett. {\bf 163A} (1992) 49.
\bibitem{Exet}
F.W. Nijhoff and H.W. Capel, in Proc. of the NATO ARW on {\em Applications
of Analytic and Geometric Methods to Nonlinear Differential Equations}, Exeter,
July 1992, ed. P. Clarkson, (Kluwer Edit. Co. , to be published).
\bibitem{AF}
A. Alekseev and L.D. Faddeev, Commun. Math. Phys. {\bf 141} (1991) 413.
\bibitem{Skly}
E.K. Sklyanin, J. Phys. {\bf A21} (1988) 2375.
\bibitem{Cher}
I.V. Cherednik, Theor. Math. Phys. {\bf 61} (1984) 35.
\bibitem{TTF}
V.O. Tarasov, L.A. Takhtajan and L.D. Faddeev, Theor. Math. Phys.
{\bf 57} (1984) 1059.
\bibitem{Bab}
O. Babelon and L. Bonora, Phys. Lett. {\bf 253B} (1991) 365.
\bibitem{Bab0}
O. Babelon, Commun. Math. Phys. {\bf 139} (1991) 619.
\bibitem{Aleks}
A. Alekseev, L.D. Faddeev, M.A. Semenov-Tian-Shanskii and A.Yu. Volkov,
Preprint CERN-TH-5981/91.
\bibitem{AFS}
A. Alekseev, L.D. Faddeev and M.A. Semenov-Tian-Shanskii, in Proc. of the
Intl. Workshop on Quantum Groups, The Euler Intl. Math. institute, Leningrad,
ed. P.P. Kulish, (Springer Lecture Notes Math. {\bf 1510}, 1991), p. 148.
\bibitem{N}
F.W. Nijhoff, {\em Integrable Quantum Mappings of KdV Type as
Canonical Transformations on Quantum Phase Space}, Preprint INS
\# 194/92.
\bibitem{NCC}
F.W. Nijhoff and H.W. Capel, to be published.
\bibitem{RSTS}
N. Yu. Reshetikhin and M.A. Semenov-Tian-Shanskii, Lett. Math. Phys.
{\bf 19} (1990) 133.
\bibitem{Mezin}
L. Mezincescu and R. Nepomechie, in {\em Quantum Groups}, eds.
T. Curtright, D. Fairlie and C. Zachos, (World Scientific, 1991),
p. 206.
\bibitem{KulSkly}
P.P. Kulish and E.K. Sklyanin, J. Phys. {\bf A24} (1991) L435.
\bibitem{KRS}
P.P. Kulish, N. Yu. Reshetikhin and E.K. Sklyanin, Lett. Math.
Phys. {\bf 5} (1981) 393.
\bibitem{KS}
P.P. Kulish and E.K. Sklyanin, Springer Lect. Notes Phys. {\bf 151}
(1982) 61.
\bibitem{KRR}
A.N. Kirillow and N.Yu. Reshetikhin, Lett. Math. Phys. {\bf 12}
(1986) 199.
\bibitem{Ols}
G.I. Olshanskii,
in {\em Topics in Representation Theory}, ed. A.A.
Kirillow, Adv. Sov. Math. {\bf 2} (1991).
\bibitem{Nepo}
L. Mezincescu and R.I. Nepomechie, Preprint CERN-TH. 6152/91.
\bibitem{Bab2}
O. Babelon, Phys. Lett. {\bf 215B} (1988) 523, ibid. {\bf 238B} (1990)
234.
\bibitem{FadSal}
L.D. Faddeev, in Proc. of the XIX ICGTMP, Salamanca, June-July 1992,
eds. J. Mateos-Guilarte, M.A. del Olmo and M. Santander (to be published in
Anales de F\'isica, Monograf\'ias, CIEMAT Publ., Spain).
\bibitem{Smirn}
F.A. Smirnov, {\em Dynamical Symmetries of Massive Integrable Models},
Preprints RIMS-772 and RIMS-838, 1991.
\bibitem{FR}
I.B. Frenkel and N.Yu. Reshetikhin, Commun. Math. Phys. {\bf 146}
(1992) 1.

\end{thebibliography}
\end{document}